\documentclass[12pt]{iopart}

\usepackage[utf8]{inputenc}
\usepackage[OT4]{fontenc}
\usepackage{graphicx} 
\usepackage{xcolor}
\usepackage{url}
\usepackage{cite}
\usepackage{tikz}
\usepackage{rotating}
\usepackage{amssymb}
\usetikzlibrary{shapes,arrows}
\usepackage{amsmath}
\usepackage{multirow}

\begin{document}

\title[CINNAMON: A hybrid approach to CPD and parameter estimation in SPT data]{CINNAMON: A hybrid approach to change point detection and parameter estimation in single-particle tracking data}

\author{Jakub Malinowski$^{1,3}$, Marcin Kostrzewa$^2$, Michał Balcerek$^1$,  Weronika Tomczuk$^1$, Janusz Szwabiński$^1$}

\address{$^1$ Hugo Steinhaus Center, Faculty of Pure and Applied Mathematics, Wrocław University of Science and Technology, Wybrzeże Stanisława Wyspiańskiego 27, 50-370 Wrocław, Poland}
\address{$^2$ Department of Artificial Intelligence, Faculty of Information and Communication Technology, Wrocław University of Science and Technology, Wybrzeże Stanisława Wyspiańskiego 27, 50-370 Wrocław, Poland}
\address{$^3$Dioscuri Centre in Topological Data Analysis, Mathematical Institute, Polish Academy of Sciences, ul. Śniadeckich 8, 00-656 Warsaw, Poland}

\ead{jakub.malinowski@pwr.edu.pl}

\vspace{10pt}
\begin{indented}
\item[]
\end{indented}

\begin{abstract}

Change point detection has become an important part of the analysis of the single-particle tracking data, as it allows one to identify moments, in which the motion patterns of observed particles undergo significant changes. The segmentation of diffusive trajectories based on those moments may provide insight into various phenomena in soft condensed matter and biological physics. In this paper, we propose CINNAMON, a hybrid approach to classifying single-particle tracking trajectories, detecting change points within them, and estimating diffusion parameters in the segments between the change points. Our method is based on a combination
of neural networks, feature-based machine learning, and statistical techniques. It has been benchmarked in the second Anomalous Diffusion Challenge. The method offers a high level of interpretability due to its analytical and feature-based components. A potential use of features from topological data analysis is also discussed.

\end{abstract}

%\maketitle

\section{Introduction}
\label{sec:intro}

%\JS{https://publishingsupport.iopscience.iop.org/journals/journal-of-physics-photonics/}

Single-particle tracking (SPT) is a class of experimental techniques and mathematical algorithms that allow small particles to be followed with nanoscale spatial localization and millisecond temporal resolution~\cite{SHE17}. It has been introduced by Perrin~\cite{PER10} and Nordlund~\cite{NOR14}, evolved in the last century, and became a popular tool that allowed quantitative analysis of particles moving inside living cells~\cite{MAN15b}. For example, SPT experiments can directly observe the diffusion of lipids and proteins in cell membranes, providing significant insight into their structure.

In a typical SPT experiment, the molecule of interest is labeled with an observable tag. The positions of the latter are tracked over a number of time steps, and then linked into a time series called trajectory. Possible tags include gold nanoparticles~\cite{SUZ05}, quantum dots~\cite{CLA13}, fluorophores~\cite{SAH10} or latex beads~\cite{CAI06}. The trajectories are then analyzed to infer, among other things, the underlying model governing the motion of particles.

Molecules with the same chemical identity can display different motion patterns as a result of interactions with the complex environment where the diffusion takes place. Potential mechanisms that impact their behavior include lipid microdomains~\cite{LIN10,EGG09}, compartmentalization by the cytoskeleton~\cite{KUS05,CLA13}, protein-protein interactions~\cite{CAI06} and inhomogeneity in the plasma membrane environment~\cite{MAS14}.

A number of methods has been already proposed for analyzing SPT data. The most common approach is based on the mean-squared displacement (MSD), which measures the deviation of particles with respect to a reference position~\cite{QIA91,MET14}.
An MSD linear in time corresponds to Brownian motion (a.k.a. normal diffusion), which describes the movement of a microscopical particle as a consequence of thermal forces. Any deviations from that linear behavior are called anomalous diffusion~\cite{KLA08}.

Although quite simple, the MSD-based analysis of trajectories is challenging because the trajectories are usually short and noisy. Consequently, several other approaches have been introduced as well. For example, the radius of gyration~\cite{SAX93}, the velocity autocorrelation function~\cite{GRE13,FUL17}, the time-dependent directional persistence of trajectories~\cite{RAU07}, the distribution of directional changes~\cite{BUR13}, the mean maximum excursion method~\cite{TEJ10}, or the fractionally integrated moving average framework~\cite{BUR15} may efficiently replace the MSD estimator for analysis purposes. 

Due to the advent of powerful GPUs and TPUs and the recent advances in machine learning (ML) algorithms, the latter has become an interesting alternative for the analysis of anomalous diffusion.  ML is already known to excel in different domains including computer vision~\cite{KRI12}, speech recognition~\cite{HIN12} and natural language processing~\cite{VAS17}. Its first applications to SPT data turned out to be promising, even though the characterization of particle motion remains very challenging~\cite{MON12, THA18, CHE17, WAG17, KOW19, MUN20a,MUN20c, JAN20, LOC20,DOS16, BO19,GRA19,GEN21,GAJ21}.

To gain some insight into the performance of various methods, a group of researchers has launched the Anomalous Diffusion (AnDi) Challenge in 2021~\cite{MUN21}. The goal of the competition was actually threefold. In addition to benchmarking existing approaches on standardized data sets, the organizers also aimed to stimulate the development of new methods and to gain new insights into anomalous diffusion. The challenge was divided into three tasks: (1) inferring the anomalous diffusion exponent from single trajectories, (2) classification of diffusion models, and (3) segmentation of trajectories. The results of the challenge clearly showed that deep learning methods based on deep neural networks significantly outperform the other ones (including feature-based machine learning), with statistical approaches being the worst~\cite{MUN21}.  

The discussion between members of various research communities sparked by the AnDi Challenge emphasized the need for a deeper understanding of biologically relevant phenomena present in SPT data. In particular, the detection of switches between different diffusive behaviors within a single trajectory requires further elaboration. These switches serve as valuable indicators for the occurrence of interactions within the system. For example, molecules may exhibit variations in diffusion coefficients  due to processes like dimerization~\cite{SAX97}, ligand binding~\cite{taraban2008ligand} or conformational changes~\cite{FRA91}. And  shifts in their mode of motion may be attributed to transient immobilization or confinement at specific scaffolding sites~\cite{SAX97,SAX94,KUS05,SIM10}. 

In response to this demand, the 2nd AnDi Challenge has been designed and launched in 2023~\cite{MUN23}. Its focus was on revealing heterogeneity rather than anomalous diffusion, but, of course, the latter was also present in the provided datasets. Similarly to the first edition of the challenge, the organizers aimed to benchmark existing methods for trajectory segmentation~\cite{REQ23,QU24,JAN21,HAN20}. Moreover, the development of new methods should be encouraged in order to detect subtle changes in diffusion properties in systems where they have been overlooked so far.

In this paper, we are going to present our contribution to the 2nd AnDi Challenge. CINNAMON (i.e. Change poInt detectioN aNd pArameter estiMation fOr aNomalous diffusion) is a hybrid method using a combination of neural networks, feature-based algorithms, and analytical methods. In the single-trajectory task within the competition, it first identifies the change points by making use of a neural network consisting of a DAIN layer for data standardization, several 1D convolutional layers, LSTM layers and an attention layer~\cite{GOO16}. Then, in each segment, the anomalous exponent is estimated with a gradient boosting method based on the features designed for the first AnDi Challenge \cite{KOW22}. For the diffusion coefficient, an estimator from \cite{LAN18} is used. In the ensemble task, another neural network is applied to identify the model which is used to generate trajectories. Its architecture is very similar to that for change point detection, as it contains the same building blocks.

The paper is structured as follows. In Sec.~\ref{sec:setup}, the competition setup of the 2nd AnDi Challenge is briefly summarized. Then, in Sec.~\ref{sec:methods}, our contribution to the challenge is presented. Results are shown and discussed in Sec.~\ref{sec:results}. Some concluding remarks may be found in Sec.~\ref{sec:conclusions}. And finally,~\ref{app:features} contains a few technical details of our method.

\section{Competition setup}
\label{sec:setup}

The details of the 2nd AnDi Challenge may be found in Ref.~\cite{MUN23}. In this section we briefly summarize them for the convenience of a reader.

\subsection{Datasets and ground truth}

For the datasets used in the challenge, we relied on the \texttt{andi-datasets} Python package \cite{MUN21} provided by the authors of the competition. It allows the generation of synthetic data representing a variety of diffusive behaviours.
All the simulations are based on fractional Brownian motion (FBM), a mathematical model that accounts for both normal and anomalous diffusion~\cite{MAN68}. FBM is a zero-mean Gaussian process $(B_H(t))_{t\geq0}$ characterized by the autocovariance function
\begin{equation}
    \langle B_H(t) B_H(s)\rangle = \frac{K}{2}\left(t^{2H} + s^{2H} - |t-s|^{2H}\right), 
\end{equation}
for any { time}  $t, s \geq 0$ and the generalized diffusion coefficient $K>0$ { with units $[K] = \frac{\mathrm{length}^2}{\mathrm{time}^{2H}}$.} The parameter $H \in (0, 1)$, often referred to as the Hurst exponent, directly relates to the anomalous diffusion exponent $\alpha$ through the equation $\alpha = 2H$. When $H = \frac{1}{2}$, the process corresponds to standard Brownian motion, with independent increments and a linear mean-squared displacement. For $H<\frac{1}{2}$, the process exhibits subdiffusion, often related to crowded environments. It corresponds to a slower than linear MSD growth. In contrast, for $H>\frac{1}{2}$, FBM is super-diffusive, which implies that motion is persistent and MSD grows faster than a linear function. 

\begin{figure}[h]
    \centering

    \includegraphics[width=\linewidth]{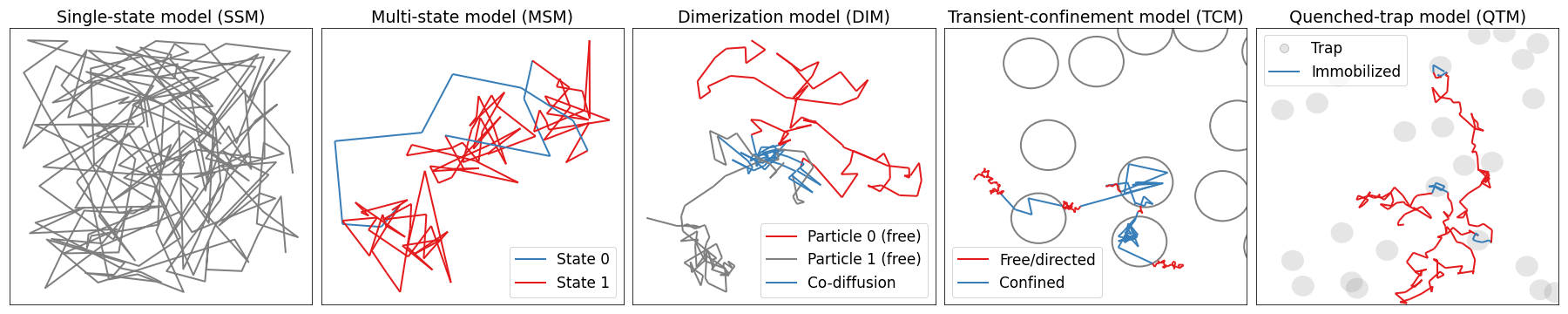}
    \caption{Physical models of motion and interactions in diffusion processes. Panels from left to right: single-state model (SSM) without changes of diffusion; multi-state model (MSM) with time-dependent changes between diffusive states (red and blue); dimerization model (DIM) with interactions between particles (red and grey) and their transient co-diffusion (blue) with a different motion; transient-confinement model (TCM) with particles diffusing inside (blue) and outside (red) compartments (circles); quenched-trap model (QTM) where a particle is transiently immobilized (blue) in specific regions (grey areas) or diffuses freely (red).}
    \label{fig:Physical_models}
\end{figure}

The generalizations of FBM considered in the challenge are based on physical models of motion and interactions. Among them, there are:
\begin{itemize}
    \item Single-state model (SSM) where the particles diffuse according to a single diffusion state. SSM serves as a baseline case, where no transitions between different motion states occur. It is often used as a control to assess false positives in methods that detect changes in the diffusive behavior. Biologically, this model can represent molecules such as lipids in the plasma membrane that diffuse freely without undergoing significant interactions or state changes.

    \item Multi-state model (MSM) where particles diffuse according to a time-dependent multi-state model of diffusion with transient changes of $K$ and/or $\alpha$. MSM describes particles that transition between different diffusion states over time. These transitions involve changes in the generalized diffusion coefficient $K$, the anomalous diffusion exponent $\alpha$, or both. The switches between states can be spontaneous or induced by interactions with the environment, leading to a heterogeneous motion pattern. In a biological context, MSM can model proteins that alter their mobility due to conformational changes, binding to ligands, or interactions with cellular structures. The duration spent in each state and the frequency of transitions determine the complexity of the resulting trajectory.
    
    \item Dimerization model (DIM) where the particles diffuse according to a 2-state model of diffusion, with transient changes of $K$ and/or $\alpha$ due to encounters with other diffusing particles. In this model, a particle moves freely until it interacts with another diffusing particle. Upon encountering its binding partner, the two particles form a temporary complex (dimer), which alters their diffusion properties. This change can manifest as a shift in the diffusion coefficient or the anomalous diffusion exponent, reflecting the fact that the dimer is typically larger and may experience different interactions with the surrounding environment. Once the binding interaction ends, the two particles separate and resume their original diffusive behavior. DIM is particularly relevant in a biological context where molecular interactions play a crucial role, such as protein dimerization, receptor-ligand binding, or transient macromolecular assemblies.
    
    \item Transient-confinement model (TCM) where particles diffuse according to a space-dependent 2-state diffusion model. In the TCM, particles diffuse freely until they enter confined regions, where their mobility is reduced due to spatial constraints. These confined regions act like compartments with osmotic boundaries that allow entry and exit. This model represents biological scenarios where molecules experience temporary entrapment, such as proteins diffusing in the plasma membrane and being transiently confined by lipid rafts or clathrin-coated pits. The confinement in TCM is not absolute; particles can escape after some time.
    \item Quenched-trap model (QTM) where particles diffuse according to a space-dependent 2-state model of diffusion. QTM represents a scenario where particles undergo intermittent immobilization by binding to static structures within the environment. Instead of being in a confined zone, particles become ``trapped'' at specific, fixed locations where their motion is temporarily halted. This is relevant for biological cases where molecules are sequestered due to interactions with elements like the cytoskeleton or scaffolding proteins. Unlike TCM, where diffusion is still possible within the confined region, in QTM, the trapped state results in a complete stop in motion until the particle is released.
\end{itemize}  
Example trajectories generated with the above models are presented in Figure \ref{fig:Physical_models}. There are no changes in the motion pattern in the SSM (left plot). All others reveal interaction-induced switches between different diffusive regimes.

\subsection{Competition design}
The 2nd AnDi Challenge was designed to evaluate methods for analyzing motion changes in single-molecule experiments and was divided into two tracks. In the first one, the participants worked with raw video data, extracting motion information directly from simulated microscopy recordings. In the second track, participants analyzed preprocessed particle trajectories, i.e., numerical values of coordinates of particles rather than images. Our team focused on the second track, dedicating our efforts to analyzing preprocessed data rather than working with raw video recordings.

\begin{figure}[h]
    \centering
    \includegraphics[width=\linewidth]{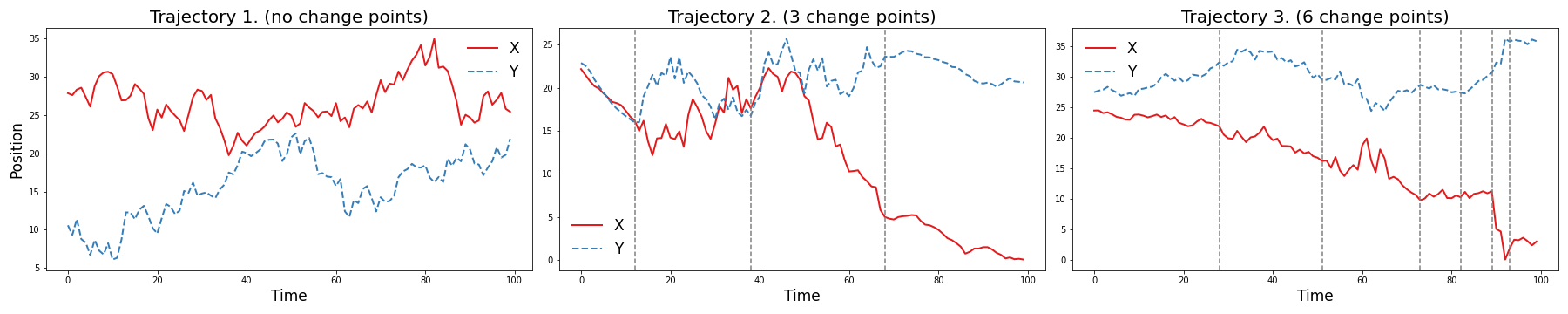}
    \caption{Exemplary particle trajectories with change points, which are indicated with dashed vertical lines. The number of change points increases from 0 (left panel), to 3 (middle panel), and 6 (right panel). Both $X$ (solid red lines) and $Y$ (dashed blue lines) coordinates of a trajectory follow the same underlying state.}
    \label{fig:change_points}
\end{figure}

In each track, participants could compete in two different tasks. The first was the ensemble task, where participants provided a statistical description of diffusion behavior across an experiment. This included identifying the underlying model, determining the number of diffusive states, and estimating the statistical properties of the diffusion coefficient $K$ and anomalous exponent $\alpha$. The second was the single-trajectory task, in which participants analyzed individual particle trajectories to detect change points marking transitions between different motion states (see Fig.~\ref{fig:change_points}). For each segment, they estimated the diffusion coefficient, the anomalous exponent, and classified the type of diffusion.

\section{Methods}
\label{sec:methods}

The method we proposed for the first AnDi Challenge was using a set of human-engineered features and the extreme gradient boosting classifier to identify diffusion modes from single trajectories~\cite{MUN21}. We deliberately decided to pick traditional machine learning over the state-of-the-art technology based on deep neural networks due to its superior interpretability. Indeed, our approach allowed us to identify factors crucial for the determination of a diffusion type in every data sample. Moreover, even though the performance of our original method was noticeably smaller than the top three deep learning solutions, we were able to significantly improve it with a further extension of the feature set~\cite{KOW22}.

Our contribution to the 2nd AnDi Challenge roots, at least to some extent, in our previous experiences. CINNAMON (Change poInt detectioN aNd pArameter estiMation fOr aNomalous diffusion) is a hybrid method using a combination of neural networks, feature-based algorithms, and analytical methods. Its details are presented in the remaining part of this section. { The implementation of the method may be found on Github~\cite{MAL24} }.

\subsection{Single-trajectory task}

\begin{figure}
    \centering
    \includegraphics[width=0.8\linewidth]{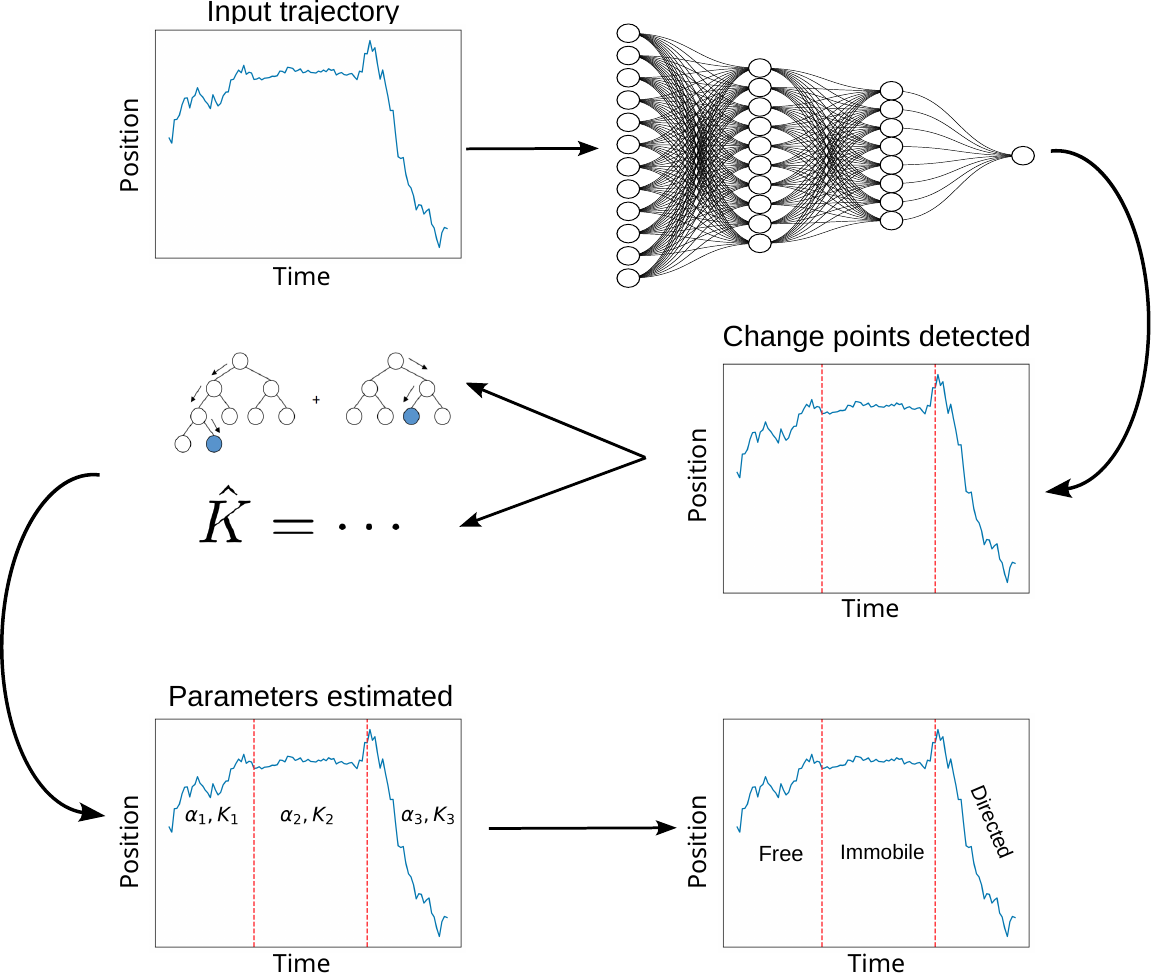}
    \caption{CINAMMON method for single-trajectory task. A neural network identifies the change points within an input trajectory. Then, in each segment, the anomalous
exponent $\alpha$ is estimated with a gradient boosting method based on the features designed for the first AnDi Challenge~\cite{KOW22}. For the generalized diffusion coefficient $K$, an estimator from~\cite{LAN18} is used (see Sec.~\ref{subsec:K}). Finally, the segments are classified based on the values of $\alpha$ and $K$.}
    \label{fig:single_task_diagram}
\end{figure}

The scheme of our method for the single-trajectory task is presented in Fig.~\ref{fig:single_task_diagram}. A neural network is used to identify change points within an input trajectory. Then, in each segment, the anomalous  exponent $\alpha$ is estimated with a gradient boosting method with the features engineered for the first AnDi challenge~\cite{MUN21,KOW22}. For the generalized diffusion coefficient $K$, an estimator from~\cite{LAN18} is used. Finally, the segments are classified according to the values of $\alpha$ and $K$. The details of each step of the procedure are described below.

\subsubsection{Change point detection}
\label{subsec:cpd}

\begin{figure}
    \centering
    \includegraphics[width=0.8\linewidth]{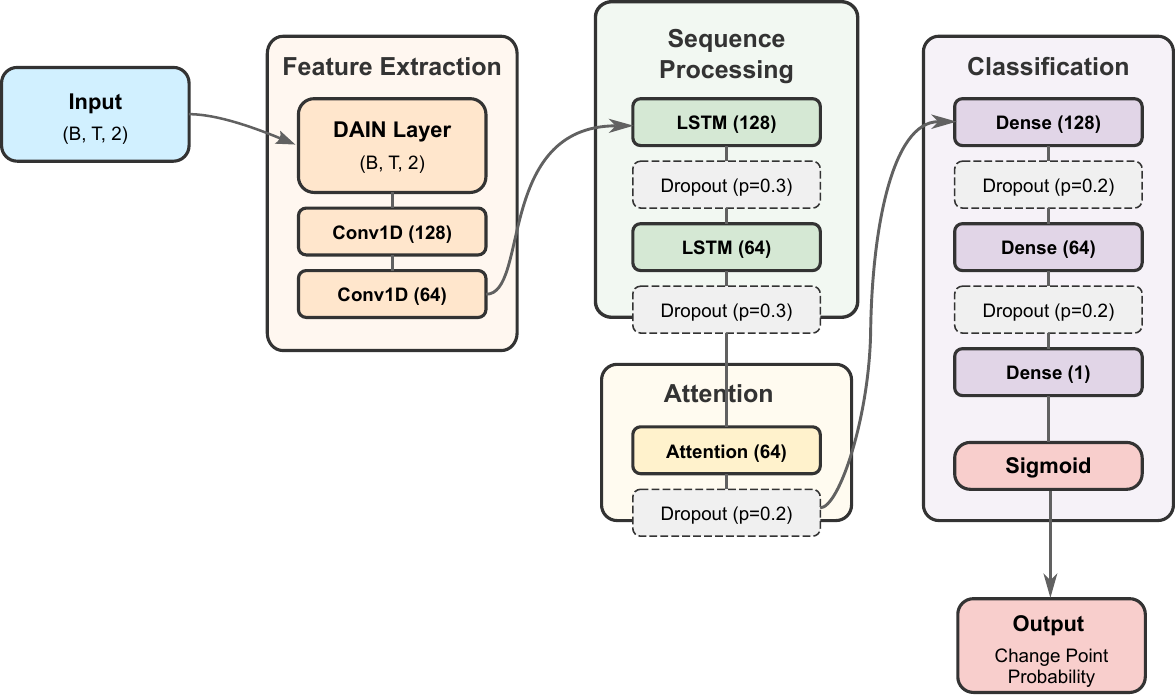}
    \caption{The architecture of neural network for change point detection. B and T on the diagram represent batch size and length of the sequence, respectively. $p$ is the dropout rate. The architecture begins with a Deep Adaptive Input Normalization
(DAIN) layer, which learns optimal normalization parameters for the time series
data. The neural
architecture then employs two 1D convolutional layers, respectively, followed by two LSTM layers  and
an attention layer with 64 units. The final prediction is made through three dense
layers of decreasing size with dropout layers included to prevent
overfitting. ReLU activation functions are
used throughout the network, except for the final layer which is followed by a sigmoid
activation to output probabilities.}
    \label{fig:nn-diagram}
\end{figure}

Our change point detection method builds upon the framework proposed by Li et al. \cite{FRY24}, employing a classifier-based approach to predict change points within trajectory subsequences. The architecture begins with a Deep Adaptive Input Normalization (DAIN) layer \cite{PAS20}, which learns optimal normalization parameters for the time series data, enhancing the model's ability to handle non-stationary sequences. The neural architecture then employs two 1D convolutional layers \cite{KIR21} with 128 and 64 feature maps, respectively, followed by two LSTM layers \cite{HOC97} containing 128 and 64 cells, and an attention layer \cite{LIU19} with 64 units. The final prediction is made through three dense layers of decreasing size (128, 64, 1 neurons), with dropout layers \cite{HIN14} included to prevent overfitting. The first two dropout layers after the LSTM layers have the probability $p$ (i.e., the dropout rate) set to $0.3$, while the subsequent dropout layers have $p = 0.2$. The complete architecture with output shapes for each layer is presented in Fig.~\ref{fig:nn-diagram}. ReLU activation functions are used throughout the network, except for the final layer which is followed by a sigmoid activation to output probabilities. The model is trained using binary cross-entropy loss.

{ The weights and biases of the neural network were initialized with the popular Glorot method~\cite{GLO10}. The model was trained over 10 epochs using the Adam optimizer~\cite{KIN14} with a learning rate equal to 0.001 and a batch size of 32. Training and validation metrics demonstrated consistent performance improvement with no signs of overfitting.}

\begin{figure}
    \centering
    \includegraphics[width=0.8\linewidth]{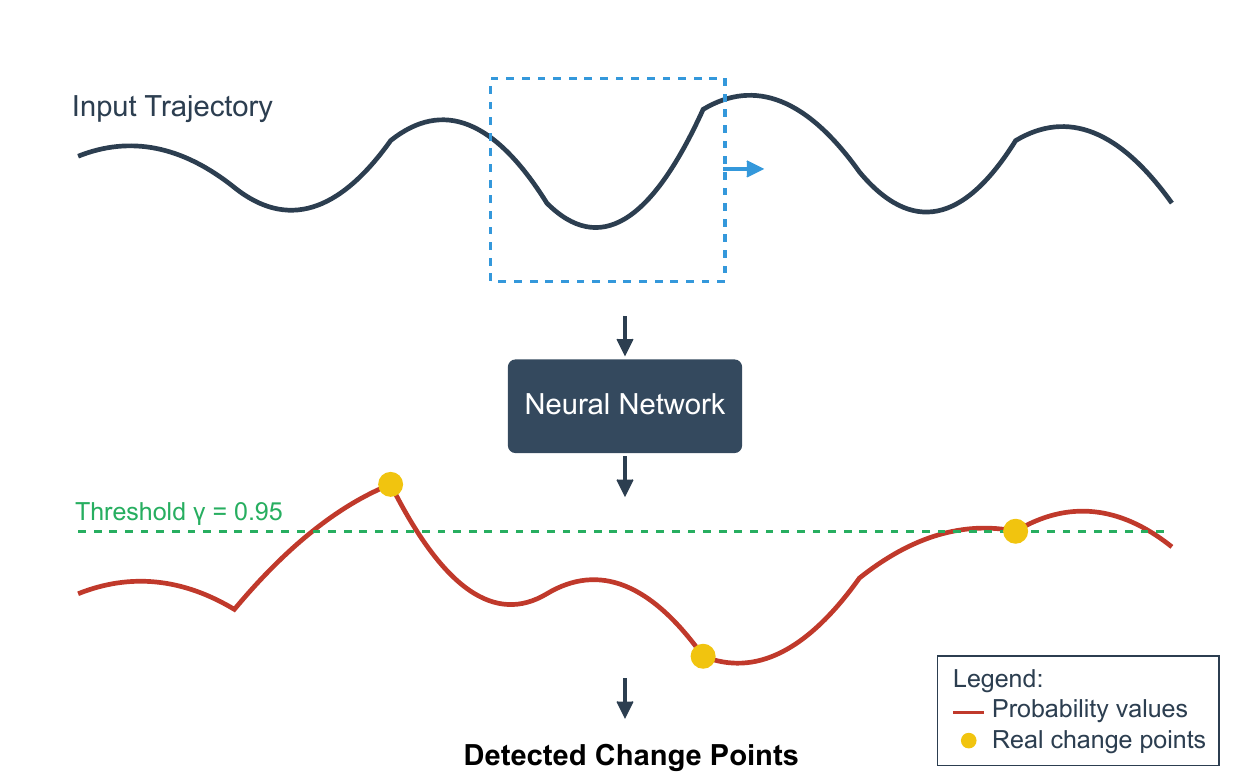}
    \caption{ Explanation of the change point detection method. The classifier shown in Fig.~\ref{fig:nn-diagram}  analyses a window of length $T=15$ sliding over a trajectory in order to generate probabilities $p_i$, which indicate the likelihood of a change point occuring. The algorithm identifies then segments where $p_i$ exceeds a threshold $\gamma=0.95$. Within each segment, the precise location of the change point is determined by finding the position where the probability reaches its maximum value. In the example presented in the diagram, the left change point would be identified correctly, the right one would be found as well but at wrong position and the method would fail to identify the point in the middle.}
    \label{fig:cpd-explained}
\end{figure}

{ For a trajectory $X = (x_1, \ldots, x_t)$, the trained classifier $\varphi: \mathbb{R}^{d\times T} \rightarrow \left[0,1\right]$ analyzes sliding windows of length $T$, generating probabilities $p_i = \varphi\left(X\left[i,i+T\right)\right)$, which indicate the likelihood of a change point occurring (see Fig.~\ref{fig:cpd-explained}). The detection algorithm identifies segments where $p_i$ exceeds a threshold $\gamma$ ($\gamma=0.95$ was used throughout this work). Within each segment, the precise location of the change point is determined by finding the position where the probability reaches its maximum value. This approach eliminates the issue of detecting consecutive change points, as only a single point from each continuous segment above the threshold is classified as a change point. 
The method assumes that change points can be detected through local patterns in the trajectory data, with the window length $T$ determining the scale of these patterns.  Experiments revealed that while classifiers trained on longer subsequences ($T = 20$) showed better classification accuracy, a window length of $T = 15$ provides the optimal balance between detection accuracy and the ability to identify closely spaced change points.}

\subsubsection{Estimation of $\alpha$}
\label{subsec:alpha}

Building on our experience from the 1st AnDi Challenge~\cite{KOW22}, we decided to use the gradient boosting method~\cite{CHE16} together with the feature set described in \ref{app:features} to estimate the anomalous exponent $\alpha$.

\begin{figure}
    \centering
    \includegraphics[width=0.7\linewidth]{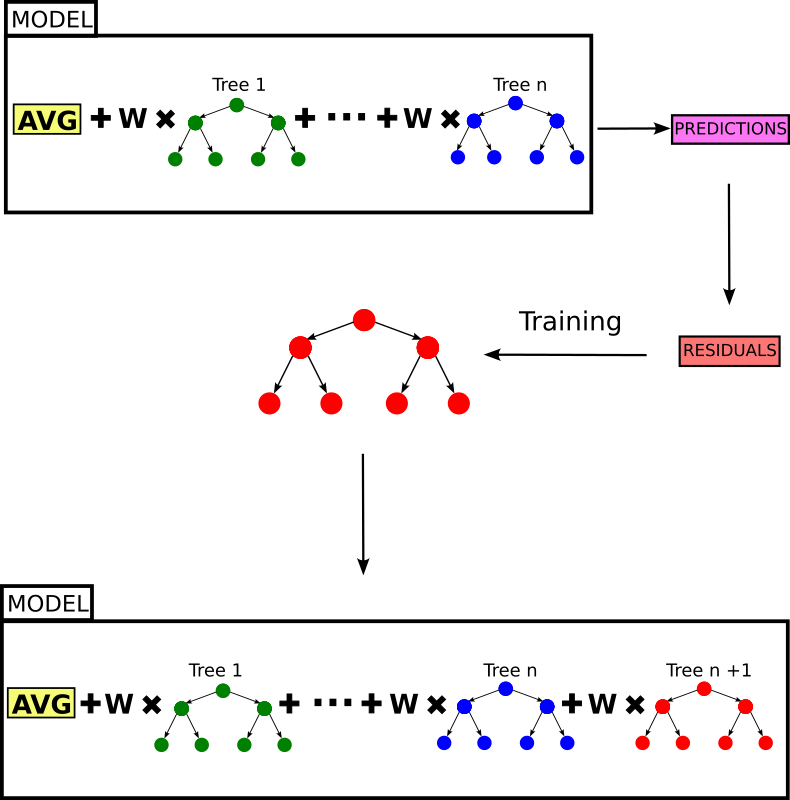}
    \caption{Construction process of the gradient boosting regressor. The algorithm begins with a naive model that simply returns the mean of training labels (called AVG in the plot). It then progressively improves prediction accuracy by sequentially adding new decision trees, each trained specifically on the errors (residuals) generated by the combined ensemble of previous trees. This iterative approach of making many small corrections to address remaining errors leads to better generalization performance and robustness compared to single decision trees. Each new tree focuses on the samples that were most challenging for the existing ensemble. The symbol $W$ in the model stands for the learning rate hyperparameter.}
    \label{fig:gb_picture}
\end{figure}

Gradient boosting is an ensemble learning method that builds a strong predictive model by sequentially training weak learners, typically decision trees~\cite{SON15}. At each iteration, a new tree is fitted to correct the errors made by the previous trees, using the gradient of the loss function as a guide (see Fig.~\ref{fig:gb_picture} for an illustration of the training process). This stepwise improvement minimizes prediction errors while maintaining flexibility in handling complex data distributions. Compared to other ensemble methods like random forest, gradient boosting tends to achieve higher predictive accuracy but may require careful tuning to avoid overfitting~\cite{FRI01}.

\subsubsection{Estimation of $K$}
\label{subsec:K}

Estimating the anomalous diffusion coefficient $K$ usually  requires a careful approach~\cite{LAN18}. Even if the experimental data exhibits a behaviour adequate to some theoretical model of anomalous diffusion, it is always disturbed by some measurement noise
\begin{equation}
    \tilde X(t) = X(t) + \xi(t),
\end{equation}
where $X(t)$ is a pure anomalous diffusion process with diffusion coefficient $K$ and exponent $\alpha$, $\xi(t)$ denotes noise, which is assumed to be independent of $X(t)$ and normally distributed with zero mean and variance $\sigma^2$ for each $t$.

The ensemble-average MSD for such a model is given by
\begin{equation}
    \langle \tilde X^2(n\Delta t)\rangle = 2K (n\Delta t)^\alpha + \sigma^2,
    \label{eq:msd}
\end{equation}
which is significantly different from the pure anomalous diffusion model for which $\sigma^2 = 0$.

The last expression may be rewritten as
\begin{equation}
        \langle \tilde X^2(n\Delta t)\rangle = 2K (n\Delta t)^\alpha + \sigma^2   
        = 2K(n\Delta t)^\alpha \left( 1 + \frac{\sigma^2}{2K(n \Delta t)^\alpha)}\right) 
\end{equation}
Taking the logarithms of both sides yields
\begin{equation}
   \label{eq:ln-msd}
   \ln \langle \tilde{X}^2(n\Delta t) \rangle = \ln (2K) + \alpha \ln (n\Delta t) + \ln \left(1 + \frac{\sigma^2}{2K (n\Delta t)^{\alpha}} \right).
\end{equation}

For sufficiently large values of $n\Delta t$, the noise term disappears, leading to a linear dependence:
\begin{equation}
    \ln \langle \tilde{X}^2(n\Delta t) \rangle \approx \ln (2K) + \alpha \ln (n\Delta t).
\end{equation}

Following the classical approach (Approach I in Ref.~\cite{LAN18}), the estimation of $K$ and $\alpha$ is performed through a least-squares fitting. The anomalous diffusion exponent $\alpha$ is then estimated as:
\begin{align}
    \hat{\alpha} = \frac{n \sum_{\tau=\tau_{\min}}^{\tau_{\max}} \ln(\tau) \ln(\widehat{MSD}(\tau)) - \sum_{\tau=\tau_{\min}}^{\tau_{\max}} \ln(\tau) \sum_{\tau=\tau_{\min}}^{\tau_{\max}} \ln(\widehat{MSD}(\tau))}
    {n \sum_{\tau=\tau_{\min}}^{\tau_{\max}} \ln^2(\tau) - \left(\sum_{\tau=\tau_{\min}}^{\tau_{\max}} \ln(\tau)\right)^2}.
\end{align}
In the last equation, the ensemble average MSD, $\langle \tilde{X}^2(n\Delta t) \rangle$, was replaced by the time-averaged MSD (TAMSD). The estimator of the latter for a sample $(x_1, x_2, x_3, \ldots, x_N)$ measured in times $\Delta t, 2\Delta t, \ldots, N\Delta t$ is defined as
\begin{equation}
 \widehat{MSD}(n\Delta t) = \frac{1}{N-n} \sum_{k=1}^{N-n} \left(x_{k+n} - x_k\right)^2,
 \label{eq:tamsd}
\end{equation}
for $n=0, 1, 2, \ldots, N-1$.
\color{black}

Once $\hat{\alpha}$ is determined, the diffusion coefficient $K$ is extracted from the intercept of the fitted regression line:
\begin{equation}
    \hat{K} = \frac{\exp(\ln \widehat{MSD}(\tau_{\min}))}{2 \tau_{\min}^{\hat{\alpha}}}.
\end{equation}

By employing the TAMSD approach, this method ensures a more robust estimation of $K$ in the presence of measurement noise compared to classical ensemble-averaged MSD methods. The choice of $\tau_{\min}$ and $\tau_{\max}$ is crucial, as discussed in \cite{LAN18}, since it defines the region where noise effects are minimized while maintaining a sufficient number of data points for a reliable estimation. In our implementation, we set \( \tau_{\min} = 1 \) and \( \tau_{\max} = 10 \) to balance statistical accuracy and noise minimization. However, a more refined selection of both \( \tau_{\min} \) and \( \tau_{\max} \) could likely improve the robustness of the estimation. { Alternatively, one could utilize non-linear regression of formula (\ref{eq:ln-msd}), with logatithm term, or its Taylor expansion, as in \cite{LAN18}. It could lead to a better $\alpha$ and $K$ estimation results, at a cost of higher computational cost. However, the trajectories available in the challenge were usually quite short, and such an approach would often lead to higher level of statistical errors.}

\subsubsection{Classification of segments}
\label{sec:classification_of_segments}

Classification of segments of the trajectories turned out to be the most challenging part of the single-trajectory task.  {We tested several approaches, including the one that we proposed for the 1st AnDi Challenge~\cite{KOW22} (referred to as the GB-Classifier~I) and finally ended up with a very simple procedure which classifies the segments based on the estimated values of $\alpha$ and $K$ (referred to as the naive method)}. Moreover, we completely neglected the ``confined'' state in the procedure (see Ref.~\cite{MUN23} for explanation). Taking it into account would make the method more complicated with practically no impact on its accuracy.

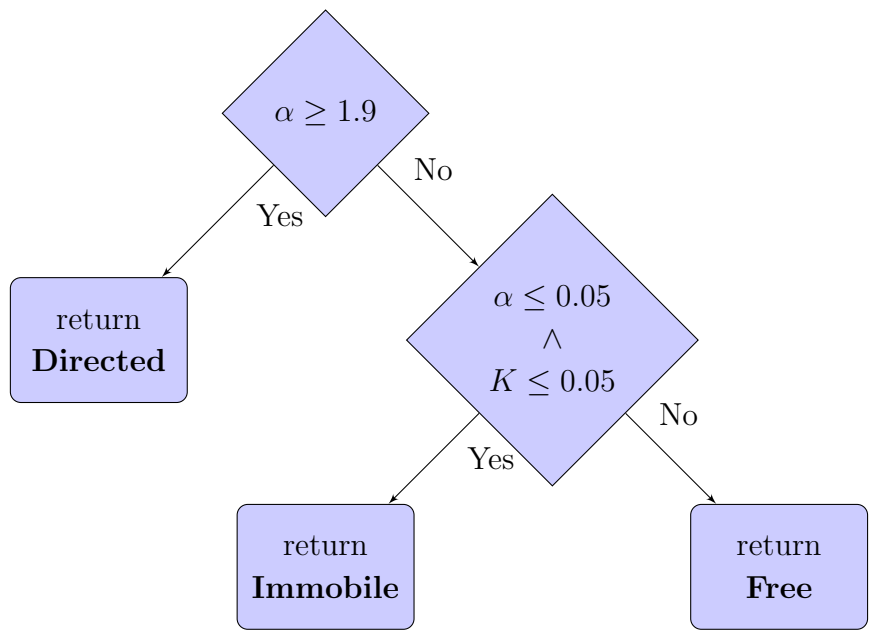
\begin{figure}
    \centering

\begin{tikzpicture}[auto, >=latex']
    % Define block styles
    \tikzstyle{block} = [draw, rectangle, minimum height=3em, minimum width=6em]
    \tikzstyle{decision} = [diamond, draw, fill=blue!20, 
                        text width=4.5em, text badly centered, node distance=3cm]
    \tikzstyle{block} = [rectangle, draw, fill=blue!20, node distance=3cm,
                        text width=5em, text centered, rounded corners, minimum height=4em]
    \tikzstyle{line} = [draw, -latex']

    % Place the nodes
    \node [decision] (decide1) {$\alpha \geq 1.9$ };
    \node [block, below of=decide1, left of=decide1] (block1) {return \textbf{Directed}};
    \node [decision, below of=decide1, right of=decide1] (block2) {$\alpha \leq 0.05$ $\land$ $K \leq 0.05$};
    \node [block, below of=block2,  right of=block2] (block4) {return \textbf{Free}};
    \node [block, below of=block2, left of=block2] (block3) {return \textbf{Immobile}};

    % Draw the edges
    \path [line] (decide1) -- node [near start] {Yes} (block1);
    \path [line] (decide1) -- node [near start] {No} (block2);
    \path [line] (block2) -- node [near start] {Yes} (block3);
    \path [line] (block2) -- node [near start] {No} (block4);

\end{tikzpicture}

    \caption{Our procedure for classification of segments. If for a given segment the estimated $\alpha$ was large, the movement of the molecule in that segment was classified as directed. If, on the other hand, both $\alpha$ and $K$ were small, the molecule was considered to be immobile. In the remaining cases, the movement of the molecule was assumed to be free.}
    \label{fig:class_seg}
\end{figure}

The naive method is shown in Fig.~\ref{fig:class_seg}. If for a given segment the estimated $\alpha$ was large, the movement of the molecule in that segment was classified as directed. If, on the other hand, both $\alpha$ and $K$ were small, we considered the molecule to be immobile. In the remaining cases, we considered the movement of the molecule to be free.

{ In Table~\ref{tab:state_score} (Sec.~\ref{sec:diffusive_states}), a brief comparison of the naive method and the GB-Classifier~I is presented. A third method, being a combination of those two, is also considered there (GB-Classifier~II). In the latter approach, a gradient boosting classifier working with the features from Appendix~\ref{app:features} was used to distinguish the ``confinement'' state from the other ones. As for the rest of the states, it was processed further with the naive method form Fig.~\ref{fig:class_seg}.} 

\subsubsection{Additional features from TDA}
\label{subsec:tda}

The engineering of the features for the 1st AnDi Challenge (see Ref.~\cite{KOW22} and \ref{app:features}) was guided by physical intuition and the statistical properties of the diffusive models under consideration. Since the models in the second challenge differ from the previous ones, we decided to further enrich the feature set by adding characteristics derived from topological data analysis (TDA)~\cite{GHR07,CAR09}. 

Euler characteristics computed for complexes generated with the Vietoris-Rips filtration~\cite{DLO23} combined with the order of points within the trajectory (points that occur later in trajectory, occur later in filtration) have been used as additional features for our regressor. 

\color{black}

\subsection{Ensemble task}

\begin{figure}[ht]
    \centering
    \includegraphics[width=0.8\linewidth]{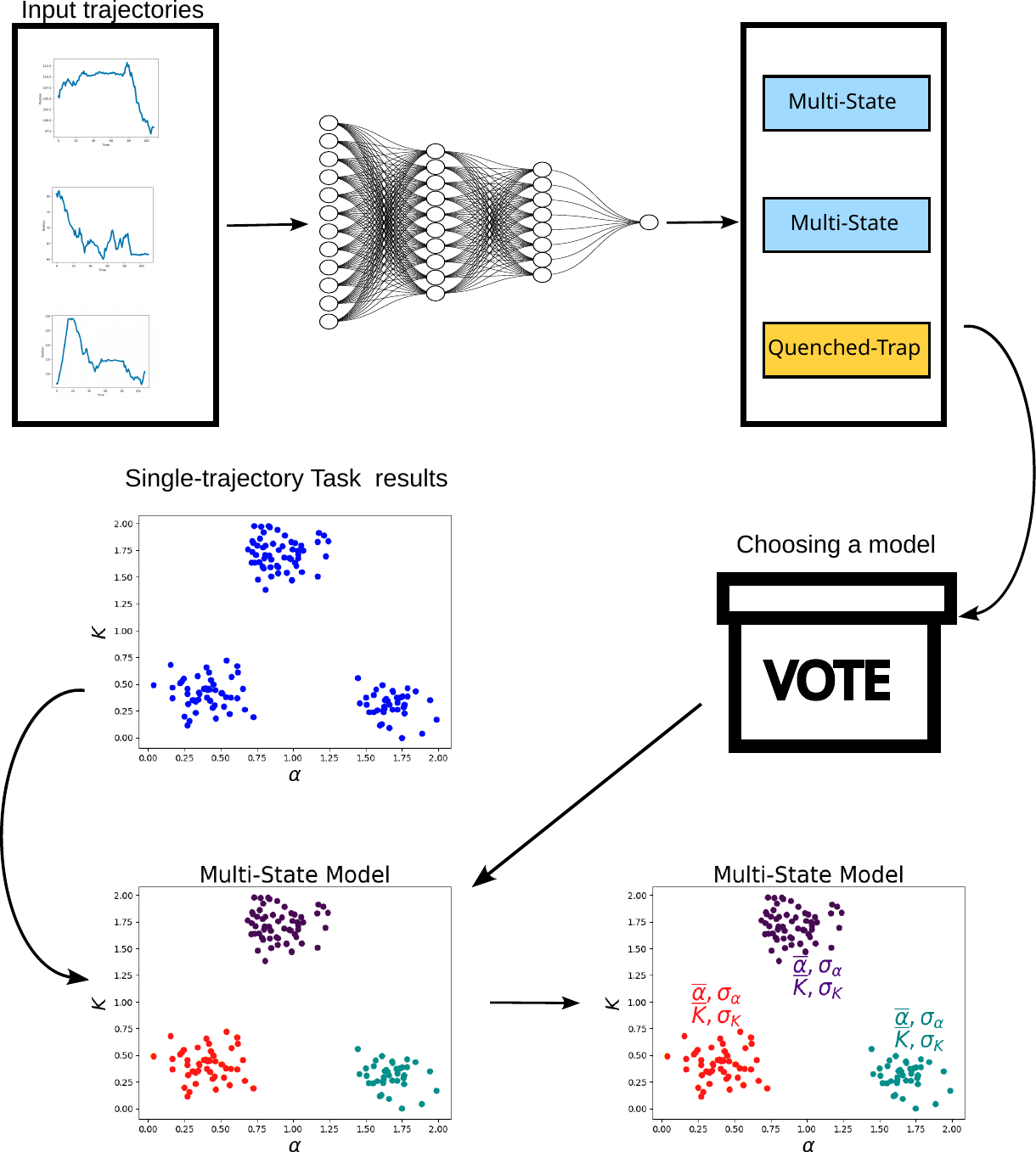}
    \caption{CINNAMON method for the ensemble task~\cite{MUN23}. First, each trajectory is classified by a neural network and then, the diffusion model used in a given experiment is identified by majority voting. Next, the results of the single-trajectory task, i.e., the pairs $(\alpha,K)$ are clustered with the $k$-means algorithm. It is simply expected that trajectories representing the same physical model should reveal similar characteristics. If the model chosen in the first step is the multi-state
model, then the number of clusters is determined based on the silhouette score. Otherwise, the model itself determines their number.
In the final step, the mean and the standard deviation of the estimates of ($\alpha$, $K$)  in each group are calculated. They are treated as the mean and the standard deviation of the parameters in the corresponding diffusive states.}

    \label{fig:ensemble_task_diagram}
\end{figure}

As already mentioned in Sec.~\ref{sec:setup}, the goal of the ensemble task in the 2nd AnDi Challenge was to provide statistical description of diffusive behavior of particles across an experiment~\cite{MUN23}. A general scheme of the CINAMMON method for this task is shown in Fig.~\ref{fig:ensemble_task_diagram}. Again, we decided to go for a hybrid solution.

In the first step of the procedure, a neural network is applied to identify the underlying model used to generate the trajectories. Its architecture is very similar to the one proposed for change point detection (see Fig.~\ref{fig:nn-diagram}). The main difference are Max\_Pooling layers~\cite{ZAF22} added after each convolution step and a Flatten layer~\cite{JEC21}. Moreover, a softmax function was used instead of a sigmoid as the last activation function. A short comparison of the architectures is given in Table~\ref{tab:single_vs_ensemble}.

After the classification of single trajectories within the experiment, its final label was determined with a majority vote, {.i.e. the most frequent label across all trajectories became the overall class of the experiment. To map between the diffusive states and their parameters $(\alpha,K)$, estimates of those parameters from the single-trajectory task were grouped together with the k-means algorithm~\cite{MAC67}. No data normalization was carried out before clusterization. For the initialization of the centroids, the greedy k-means++ algorithms~\cite{GRU23} was used. It was carried out only once for each clusterization. The number of groups was specified by the identified type of the model in most of the cases. For single-state models, only one cluster was considered. For all other models except the multi-state one, the number of clusters was set to 2. Since the multi-state model does not imply any particular number of clusters, in this case we performed the grouping sequentially for the number of clusters ranging from 2 to $2+N$, where $N$ is the number of $(\alpha,K)$ pairs. For each clustering attempt, the average silhouette score was calculated and the grouping with its highest value was selected. No specific remediation strategy was implemented for cases with low clustering  performance. 

Finally, the mean and standard deviations of the estimates of $(\alpha , K)$ in each group were calculated. The obtained values can be treated as mean and standard deviation of the parameters in the corresponding diffusive states.}

\begin{table}[ht]
    \centering
    \begin{tabular}{|c|c|c|}
    \hline
     Layer     & Single-trajectory task &  Ensemble task   \\ \hline\hline
     DAIN      &  $(B, T, 2)$   &  $(B, 150, 2)$ \\ \hline
     Conv1D    &  $(B, T, 128)$  & $(B, 148, 128)$\\ \hline
     Max\_Pooling1D & --- &  $(B, 49, 128)$ \\ \hline
     Conv1D    &  $(B, T, 64)$   & $(B, 47, 64)$ \\ \hline
     Max\_Pooling1D & --- &  $(B, 23, 64)$ \\ \hline
     LSTM      &  $(B, T, 128)$  & $(B, 23, 256)$ \\ \hline
     Dropout   &  $(B, T, 128)$  &  $(B, 23, 256)$ \\ \hline
     LSTM      &  $(B, T, 64)$   &  $(B, 23, 128)$ \\ \hline
     Dropout   &  $(B, T, 64)$   &  $(B, 23, 128)$\\ \hline
     Attention &  $(B, 64)$      &  $(B, 128)$ \\ \hline
     Dropout   &  $(B, 64)$      &  $(B, 128)$ \\ \hline
     Flatten   & --- &    $(B, 128)$ \\ \hline
     Dense    &  $(B, 128)$     & $(B, 128)$\\ \hline
     Dropout  &  $(B, 128)$     & $(B, 128)$\\  \hline
     Dense    &  $(B, 64)$      & $(B, 64)$\\  \hline
     Dropout  &  $(B, 64)$      & $(B, 64)$\\  \hline
     Dense     &  $(B, 1)$       & $(B, 5)$\\  \hline
    \end{tabular}
    \caption{Comparison of the neural network architectures used in the single-trajectory (see Fig.~\ref{fig:nn-diagram}) and the ensemble tasks. The biggest difference between the architectures are the Max\_Pooling layers added after every convolution step in the ensemble task, as well as the additional Flatten layer. The shapes of the layers differ as well. Another difference, not shown in the table, is the last activation function -- in the ensemble task a softmax function was used instead of a sigmoid. $B$ is the size of each data batch {($B=32$ was used throughout this work)}, $T$ is the length of the trajectory. }
    \label{tab:single_vs_ensemble}
\end{table}

\section{Results}
\label{sec:results}

Most of the results presented in this section are based on the benchmark dataset provided by the organizers of the 2nd AnDi Challenge \cite{MUN24}. This dataset covers a wide range of biological and physical scenarios, such as multistate diffusion, trapping, confinement, and dimerization, with experiments designed to explore diverse diffusive parameters. This allows for a thorough evaluation of algorithms for change point detection, classification, and characterization. The results of the final round of the AnDi Challenge 2 are presented in Tables \ref{tab:final_single}-\ref{tab:final_ensemble}. Our team (called HSC AI) placed 13th in the single-trajectory task and 7th in the ensemble task.

\begin{sidewaystable}[htp]
\centering
\vspace{40em}
\begin{tabular}{|c|c|c|c|c|c|c|c|c|c|c|c|c|c|}
\hline
Global & Team & RMSE & JSC & MAE & MSLE & F1 (diff. & RMSE & JSC & MAE & MSLE & F1 & MRR \\
rank & & (CP) & (CP) & (alpha) & (K) & type) & rank & rank & rank & rank & rank & \\
\hline
1 & UCL SAM & 1.639 & 0.703 & 0.175 & 0.015 & 0.968 & 1 & 1 & 1 & 1 & 1 & 1.000 \\
\hline
2 & SPT-HIT & 1.693 & 0.650 & 0.217 & 0.022 & 0.915 & 3 & 13 & 8 & 2 & 2 & 0.358 \\
\hline
3 & HNU & 1.658 & 0.482 & 0.178 & 0.060 & 0.871 & 2 & 10 & 2 & 10 & 10 & 0.260 \\
\hline
4 & M3 & 1.738 & 0.649 & 0.184 & 0.024 & 0.652 & 5 & 4 & 3 & 4 & 15 & 0.220 \\
\hline
5 & bjyong & 1.896 & 0.664 & 0.211 & 0.252 & 0.879 & 7 & 2 & 6 & 11 & 9 & 0.202 \\
\hline
6 & SU-FIONA & 2.426 & 0.579 & 0.194 & 0.024 & 0.885 & 10 & 6 & 4 & 3 & 7 & 0.199 \\
\hline
7 & Unfriendly AI & 1.709 & 0.632 & 0.241 & 0.042 & 0.903 & 4 & 5 & 9 & 8 & 4 & 0.187 \\
\hline
8 & KCL & 1.803 & 0.533 & 0.214 & 0.030 & 0.882 & 6 & 8 & 7 & 6 & 8 & 0.145 \\
\hline
9 & EmetBrown & 4.439 & 0.084 & 0.309 & 0.056 & 0.910 & 17 & 17 & 12 & 9 & 3 & 0.129 \\
\hline
10 & BIOMED-UCA & 2.138 & 0.570 & 0.275 & 0.026 & 0.859 & 9 & 7 & 10 & 5 & 12 & 0.127 \\
\hline
11 & Nanoninjas & 3.677 & 0.246 & 0.203 & 1.870 & 0.899 & 12 & 12 & 5 & 16 & 5 & 0.126 \\
\hline
12 & KNU-ON & 2.659 & 0.488 & 0.307 & 0.031 & 0.756 & 11 & 9 & 11 & 7 & 14 & 0.101 \\
\hline
13 & HSC AI & 4.025 & 0.193 & 0.393 & 1.402 & 0.896 & 15 & 15 & 15 & 15 & 6 & 0.087 \\
\hline
14 & Alntgonnawork & 1.994 & 0.447 & 0.563 & 0.298 & 0.545 & 8 & 11 & 17 & 13 & 16 & 0.083 \\
\hline
15 & ICSO UPV & 4.056 & 0.211 & 0.380 & 5.255 & 0.861 & 16 & 13 & 14 & 17 & 11 & 0.072 \\
\hline
16 & D.AnDi & 3.879 & 0.204 & 0.472 & 0.275 & 0.354 & 14 & 14 & 16 & 12 & 17 & 0.070 \\
\hline
17 & DeepSPT & 4.894 & 0.186 & 0.336 & 0.424 & 0.822 & 18 & 16 & 13 & 14 & 13 & 0.069 \\
\hline
18 & far\_naz & 3.829 & 0.023 & 1.476 & 94.564 & 0.293 & 13 & 18 & 18 & 18 & 18 & 0.060 \\
\hline
\end{tabular}
\caption{Final results of the 2nd AnDi Challenge - single-trajectory task. Our team (HSC AI) is placed 13th. See Ref.~\cite{MUN23} for the explanation of all measures used in the table.}
\label{tab:final_single}
\end{sidewaystable}

\begin{table}[h]
\centering
\begin{tabular}{|c|c|c|c|c|c|c|c|}
\hline
Global rank & Team & W1 (alpha) & W1 (K) & alpha rank & k rank & MRR \\
\hline
1 & UCL SAM & 0.138 & 0.058 & 1 & 4 & 0.250 \\
\hline
2 & DeepSPT & 0.267 & 0.050 & 9 & 1 & 0.222 \\
\hline
3 & Nanoninjas & 0.192 & 0.051 & 3 & 2 & 0.167 \\
\hline
4 & bjyong & 0.188 & 0.084 & 2 & 7 & 0.129 \\
\hline
5 & SPT-HIT & 0.231 & 0.057 & 6 & 3 & 0.100 \\
\hline
6 & ICSO UPV & 0.218 & 0.067 & 4 & 5 & 0.090 \\
\hline
7 & HSC AI & 0.230 & 0.256 & 5 & 8 & 0.065 \\
\hline
8 & Unfriendly AI & 0.238 & 0.072 & 7 & 6 & 0.062 \\
\hline
9 & EmetBrown & 0.259 & 0.622 & 8 & 11 & 0.043 \\
\hline
10 & BIOMED-UCA & 0.275 & 0.534 & 10 & 9 & 0.042 \\
\hline
11 & KCL & 0.448 & 0.593 & 11 & 10 & 0.038 \\
\hline
\end{tabular}
\caption{Final results of the 2nd AnDi Challenge - ensemble task. Our team (HSC~AI) is placed 7th. See Ref.~\cite{MUN23} for the explanation of all measures used in the table.}
\label{tab:final_ensemble}
\end{table}

\subsection{Change point detection}

\begin{figure}
    \centering
    \includegraphics[width=0.98\linewidth]{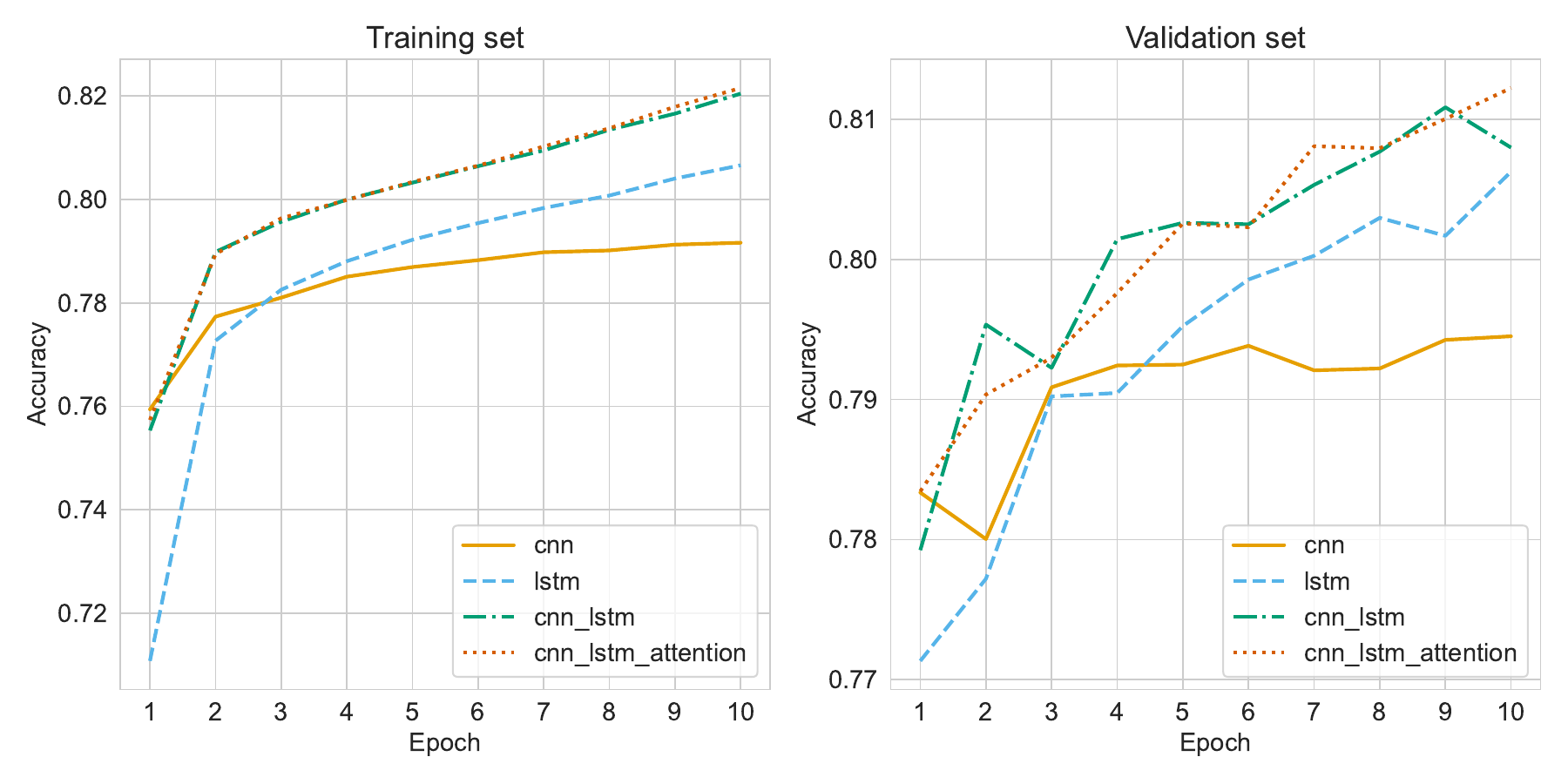}
    \caption{Training and validation accuracy for different variants of the neural network classifier. The full model, described in Sec.~\ref{subsec:cpd}, is compared with three simplified versions (CNN+LSTM, CNN-only, and LSTM-only).\label{fig:accuracy-results}}
\end{figure}

\begin{figure}
    \centering
    \includegraphics[width=0.98\linewidth]{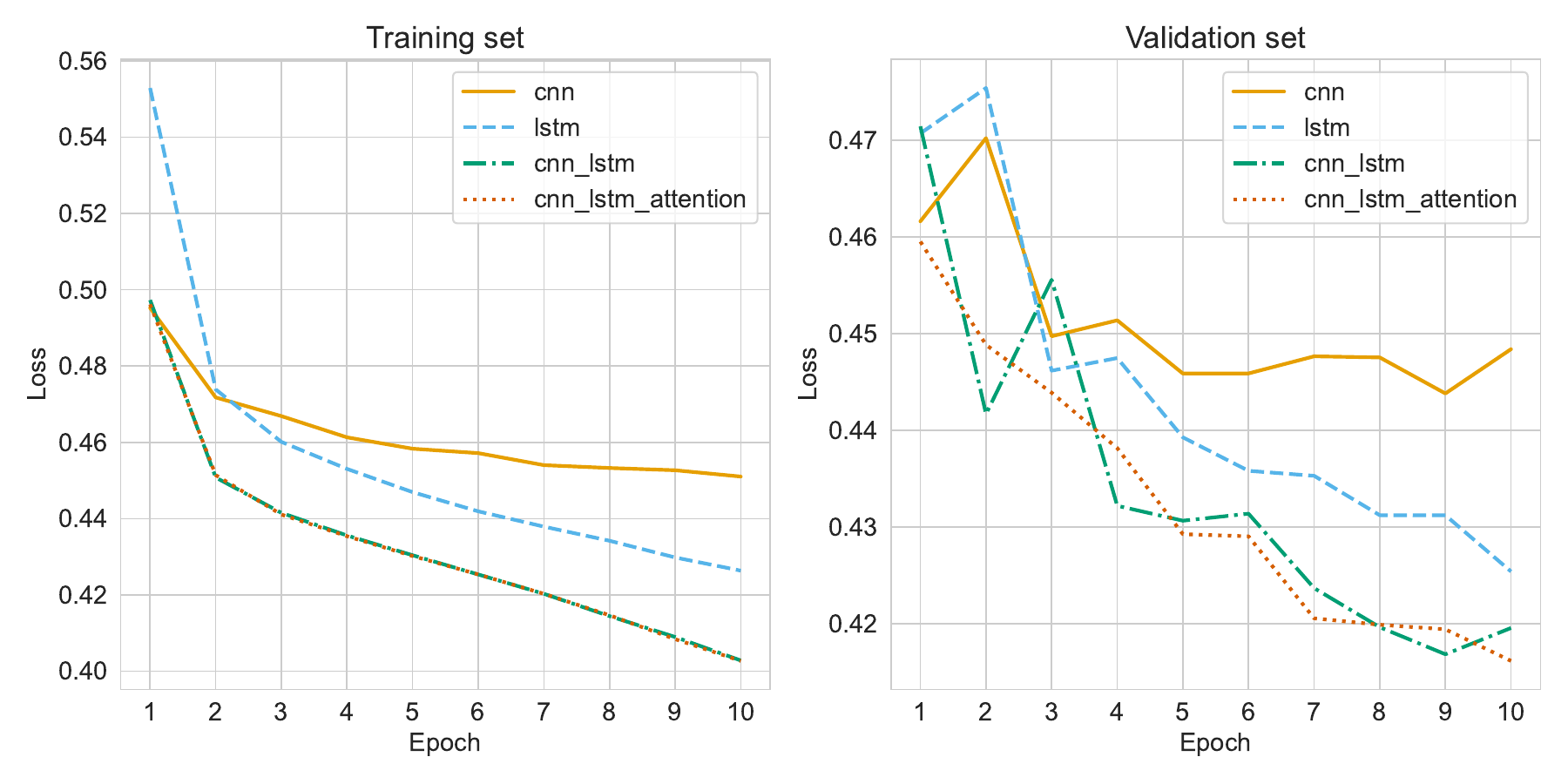}
    \caption{Training and validation loss for different variants of the neural network classifier. The full model, described in Sec.~\ref{subsec:cpd}, is compared with three simplified versions (CNN+LSTM, CNN-only, and LSTM-only).\label{fig:loss-results}}
\end{figure}

The training and validation loss and accuracy values are presented in Figure \ref{fig:loss-results} and Figure \ref{fig:accuracy-results}, respectively. Four neural network classifier architectures were compared: the full model described in Sec.~\ref{subsec:cpd}, as well as three simplified variants (CNN+LSTM, CNN-only, and LSTM-only). All architectures incorporate the DAIN layer for preprocessing. { The early plateau in the CNN-only model demonstrates that adding LSTM components significantly improves model performance, as the CNN+LSTM and the full models continue to decrease their loss throughout the 10 epochs. This confirms our hypothesis that capturing temporal
dependencies is crucial for this task. Regarding the attention mechanism, while its contribution appears
modest in the loss curves, accuracy metrics show that the full model achieves slightly better performance than CNN+LSTM (0.8140 accuracy vs. 0.8087). Even such a small improvement can be considered as valuable in the context of change-point
detection. Additionally, the attention module adds minimal computational overhead to the model (194
640 vs. 190 415 parameters).}

The results of the single-trajectory task for the benchmark dataset are presented in Table \ref{tab:benchmar_single}. With the exception of experiments 6 and 8, where this metric is equal to 0, root-mean-square error (RMSE) is at a similar level in all experiments. A greater variety of results is seen for the Jaccard similarity coefficient (JSC). No clear relationship is apparent between the model of the experiment and the effectiveness of change point detection.

\begin{table}[thp]
\centering
\begin{tabular}{|c|c|c|c|c|c|c|}
\hline
Exp & num\_trajs & RMSE (CP) & JSC (CP) & MSLE (K) & MAE (alpha) & F1 (diff. type) \\
\hline
1 & 1280 & 4.512061 & 0.012563 & 2.585260 & 0.567237 & 0.979381 \\
\hline
2 & 8983 & 4.737913 & 0.074069 & 0.540352 & 0.814740 & 0.997437 \\
\hline
3 & 1348 & 3.655631 & 0.384653 & 2.435835 & 0.328328 & 0.852941 \\
\hline
4 & 1192 & 4.524786 & 0.257439 & 2.581110 & 0.372193 & 0.601671 \\
\hline
5 & 6485 & 4.582201 & 0.155559 & 0.561066 & 0.378303 & 0.993404 \\
\hline
6 & 1172 & 0.000000 & 0.452656 & 3.122132 & 0.507152 & 0.973729 \\
\hline
7 & 1222 & 4.962651 & 0.058971 & 2.621429 & 0.471954 & 0.979050 \\
\hline
8 & 1524 & 0.000000 & 0.958172 & 4.888330 & 0.483815 & 0.951444 \\
\hline
9 & 1857 & 4.602225 & 0.136440 & 1.082408 & 1.052998 & 0.213957 \\
\hline
\end{tabular}
\caption{Results in the single-trajectory task for the benchmark dataset. No clear relationship is apparent between the
model of the experiment and the effectiveness of change point detection.\label{tab:benchmar_single}}
\end{table}

\subsection{Parameters estimation}

\begin{figure}
    \centering
    \includegraphics[width=0.7\linewidth]{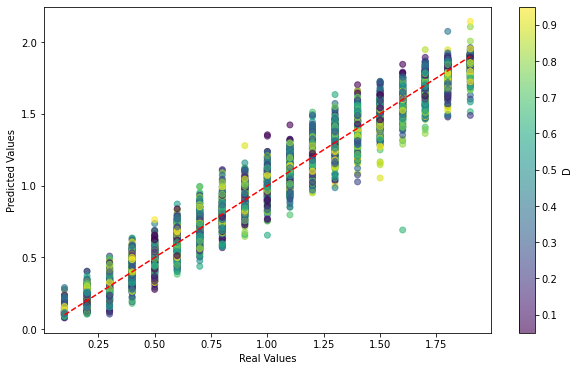}
    \caption{Prediction of $\alpha$ with the gradient boosting regressor and the feature set from the 1st AnDi Challenge. {The error metrics for these estimates are presented in Table \ref{tab:estim_results_alpha}.}\label{fig:alpha_pred}}
\end{figure}

\begin{figure}
    \centering
    \includegraphics[width=0.7\linewidth]{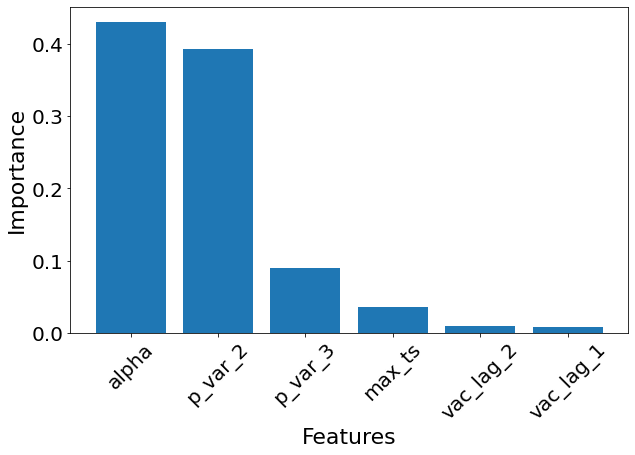}
    \caption{Importance of the six most important features for estimating $\alpha$ using gradient boosting.  \texttt{alpha} is the standard estimation of the anomalous exponent, based on fitting the empirical TAMSD to the model~(\ref{eq:msd_th}). \texttt{p\_var\_2} and \texttt{p\_var\_3} correspond to exponents $\gamma^p$ fitted to the empirical p-variation (Eq.~\ref{eq:pvar}) with $p=2$ and $p=3$, respectively. \texttt{max\_ts} is the mean maximal excursion (Eq.~(\ref{eq:ts})). And finally, \texttt{vac\_lag\_2} and \texttt{vac\_lag\_1} are the velocity autocorrelation functions at lags 2 and 1 (Eq.~(\ref{eq:vac})). Impact of the remaining features on estimation is minimal.  The importances reflect how often the features are used to split the data across all trees.\label{fig:importance} }
\end{figure}

The results of the estimation of parameter $\alpha$ using the features of the first AnDi challenge are presented in Fig.~\ref{fig:alpha_pred}. {They were obtained for trajectories of length 100.} 
The original feature set was created to solve the problem of distinguishing between different types of anomalous diffusion. In case of regression of $\alpha$, only some of them turned out to be useful. The significance of the most important ones is shown in the Fig.~\ref{fig:importance}.
The influence of the other features on the estimations is negligible. The most important features (sorted by significance) are as follows:
\begin{itemize}
    \item anomalous exponent estimation based on fitting the empirical TAMSD,
    \item statistics based on $p$-variation (for $p=2,3$),
    \item mean maximal excursion,
    \item empirical velocity autocorrelation functions.
\end{itemize}
For details regarding the features, please refer to~\ref{app:features}.

\begin{figure}
    \centering
    \includegraphics[width=0.7\linewidth]{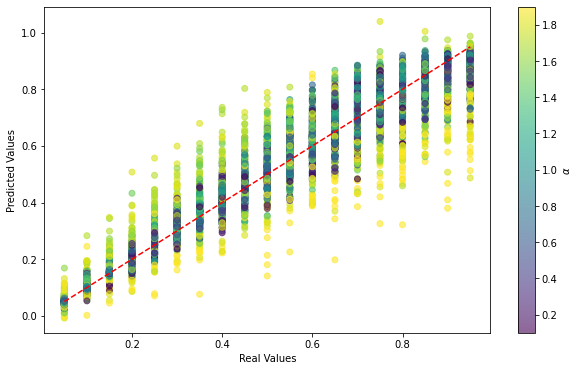}
    \caption{Prediction of $K$ with the method from Sec.~\ref{subsec:K}.  {The error metrics for these estimates are presented in Table \ref{tab:estim_results_D}.}\label{fig:D_pred}}
\end{figure}

The results for  the generalized diffusion coefficient $K$ are presented in Fig.~\ref{fig:D_pred}. We see that for both $\alpha$ and $K$, the accuracy of the prediction decreases with the increasing values of the parameters.
It is worth to mention that we tried to improve the estimation of $K$ by making use of the gradient boosting regressor we have already used for $\alpha$. However, the predictions were practically the same. Although quite suprising at first glance, this outcome can be easily explained. The feature set presented in~\ref{app:features} contains already the estimator of $K$ described in Sec.~\ref{subsec:K}. And, after building the regressor for $K$ it turned out that this estimator is the only important feature. The remaining ones can be neglected. This is the reason why we observed practically no difference between the methods. It also indicate that the feature set may be not appriopriate for the problem at hand.

Adding topological descriptors to the set of characteristics did not improve the results either. Their effect on the estimation was negligible. { There are a few potential reasons for that outcome. First of all, since all of the diffusion models were based on the fractional Brownian motion, they might not generate trajectories with significant, persistent topological differences that the Euler characteristic can effectively distinguish. Secondly, the filtration we have chosen causes the initial parts of the trajectory to have a greater impact on the topological descriptor values than the final parts. As there is no reason to favor the initial points of the trajectory, this phenomenon may have had a negative impact on the quality of the proposed descriptor.} More details regarding the performance of different approaches can be seen in Tables~\ref{tab:estim_results_alpha}-\ref{tab:estim_results_D}.

\begin{table}[htp]
    \centering
    \begin{tabular}{|c|c|c|c|}
    \hline
          & R2 & MSE & MAE  \\ \hline
          AnDi 1 features& 0.973 & 0.008 & 0.069 \\ \hline
         TDA + AnDi 1 features & 0.973 & 0.008 & 0.069\\ \hline
    \end{tabular}
    \caption{Estimation results for the anomalous exponent $\alpha$. Two variants of the regressors are compared. The one was trained only with the features described in~\ref{app:features} (refered to as AnDi 1 features). In case of the other, the feature set was extended with the characteristics from TDA (see Sec.~\ref{subsec:tda}).\label{tab:estim_results_alpha}}
\end{table}

\begin{table}[htp]
    \centering
    \begin{tabular}{|c|c|c|c|}
    \hline
          & R2 & MSE & MAE  \\ \hline
          Estimator & 0.864 & 0.010 & 0.062  \\ \hline
          AnDi 1 features& 0.923 & 0.006 & 0.051  \\ \hline
         TDA + AnDi 1 features & 0.923 & 0.006 & 0.051 \\ \hline
    \end{tabular}
    \caption{Estimation results for the generalized diffusion coefficient $K$. The analytical method from Sec.~\ref{subsec:K} is compared with two variants of the regressors. The one was trained with the features described in~\ref{app:features} (refered to as AnDi 1 features). In case of the other, the feature set was extended with the characteristics from TDA (see Sec.~\ref{subsec:tda}).\label{tab:estim_results_D}}
\end{table}

\subsection{Diffusive states}
\label{sec:diffusive_states}
Despite of the simplicity of our method we obtained high $F_1$-score on the benchmark dataset (Table \ref{tab:benchmar_single}).
For most of the experiments, the index was higher than 0.95. Worse results were obtained for the Immobile Traps experiments ( number 3 and 9) and the Confinement experiment (number 4). In these cases, the $F_1$-score was 0.85, 0.21 and 0.60, respectively. Worse results in case of Confinement model based experiment was expected, because our method does not take the confinement state into account (see Sec.~\ref{sec:classification_of_segments})

{One can of course ask the question how the predictive power of our method will change after taking that state into consideration. A comparison of our naive method with two simple feature-based classifiers is presented in Table~\ref{tab:state_score}. We see that the naive method indeed outperfoms the classifiers, which may indicate that the feature set designed for the previous edition of the AnDi Challenge may require far-reaching adjustments for the new type of data.}

\begin{table}[]
    \centering
    \begin{tabular}{|c|c|c|c|}
         \hline
         &Naive & GB-Classifier I & GB-Classifier II \\ \hline
         AVG & 0.838 &	0.812 &	0.808 \\ \hline
         Weighted AVG &0.906 &	0.835 &	0.837 \\ \hline
    \end{tabular}
    \caption{ F-1 score values for the classification of diffusive states, averaged over all experiments. Weighted averages are given in the second row (weights are the number of trajectories in each experiment). The definition of the classification methods is given in Sec.~\ref{sec:classification_of_segments}. The ``naive'' one was used throught this paper.}
    \label{tab:state_score}
\end{table}

\subsection{Ensemble Task}

The outcomes for each benchmark experiment are shown in Table \ref{tab:end_benchmark_general}. It should be emphasized here that these results  are grounded in parameter estimates for $\alpha$ and $K$ derived from the single-trajectory task. The dependency of the ensemble task on the prior one  is particularly noticeable in the estimation of the $\alpha$ parameter, with the best results for this parameter achieved in experiments 2, 3, and 4. The relationship between the results of the two tasks is more intricate for parameter $K$. In experiments 2 and 4, the $K$ parameter results are moderately successful for the single-trajectory task but substantially poor for the ensemble task, whereas Experiment 5 yields good results for both tasks.

\begin{table}[h]
\centering
\begin{tabular}{|c|c|c|}
\hline
Exp & W1 (K) & W1 (alpha) \\
\hline
1 & 0.915170 & 0.531545 \\
\hline
2 & 0.819725 & 0.782572 \\
\hline
3 & 0.502521 & 0.247958 \\
\hline
4 & 0.063257 & 0.146702 \\
\hline
5 & 0.376822 & 0.190814 \\
\hline
6 & 0.291466 & 0.304549 \\
\hline
7 & 0.754769 & 0.289931 \\
\hline
8 & 0.486562 & 0.395450 \\
\hline
9 & 1.158701 & 0.722865 \\
\hline
\end{tabular}
\caption{Results for benchmark experiments in the ensemble task. $W1$ stands for the Wasserstein distance (please refer to Ref.~\cite{MUN23} for further details).\label{tab:end_benchmark_general}}
\end{table}

%\JM{Uwagi do uwagi nr. 5 Recenzenta 1
%\begin{itemize}
%    \item Klasyfikacja modelu eksperymentu nie była bezpośrednio oceniana w AnDi Challenge 2
%    \item Z tego powodu nie skupialiśmy na tym zagadnieniu i nie próbowaliśmy polepszyć naszej metody
%    \item Jedną z wad naszej metody jest to, że klasyfikacje robimy wpierw dla każedj trajektorii osobno. W takim ustawieniu wydaje się trudno wykryć demerization, bo nie widzimy zderzeń cząstek. Podobne komplikacje dotyczą pozostałych modeli eksperymentu
%\end{itemize}}

Our method of recognizing the experiment model failed. Only one experiment from the benchmark was correctly classified (Table \ref{tab:model_class}). There is a clear bias in the prediction towards the single-state model. In many cases, misclassification of the model led to poor selection of the number of states.

\begin{table}[th]
    \centering
    \begin{tabular}{|c|c|c|}
         \hline
         & Real & Predicted  \\ \hline
         Exp. 1 & Multi-state & Confinement \\ \hline
         Exp. 2 & Dimerization & Dimerization \\ \hline
         Exp. 3 & Immobile traps & Single State \\ \hline
         Exp. 4 & Confinement & Dimerization \\ \hline
         Exp. 5 & Dimerization & Single State \\ \hline
         Exp. 6 & Dimerization & Single State \\ \hline
         Exp. 7 & Multi-state & Single State \\ \hline
         Exp. 8 & Multi-state & Single State \\ \hline
         Exp. 9 & Immobile traps & Dimerization \\ \hline
    \end{tabular}
    \caption{Results for classification of physical models in the benchmark dataset.}
    \label{tab:model_class}
\end{table}

To test how much impact these misclassifications had on the final results of our method, we replaced the predicted physical models with the real ones. The results are shown in Table \ref{tab:end_benchmark_general_model}. As can be seen, this change did not significantly improve the results. Only in the case of experiment 3, a noticeable improvement is visible.

\begin{table}[th]
\centering
\begin{tabular}{|c|c|c|}
\hline
Exp & W1 (K) & W1 (alpha) \\
\hline
1 & 0.929019 & 0.531413 \\
\hline
2 & 0.819689 & 0.782573 \\
\hline
3 & 0.339198 & 0.164255 \\
\hline
4 & 0.063257 & 0.146702 \\
\hline
5 & 0.259741 & 0.159874 \\
\hline
6 & 0.339915 & 0.341425 \\
\hline
7 & 0.736490 & 0.283159 \\
\hline
8 & 0.554527 & 0.398018 \\
\hline
9 & 1.159196 & 0.722828 \\
\hline
\end{tabular}
\caption{Results for benchmark experiments in the ensemble task when the information regarding real physical models is used. For all but the experiment 3, this additional information did not significantly improve the results.}
\label{tab:end_benchmark_general_model}
\end{table}

\section{Conclusions}
\label{sec:conclusions}

In this paper, a hybrid method for change point detection and parameter estimation for anomalous diffusion has been presented. The method is a mixture  of deep-learning algorithms, feature-based regression and analytical estimation. Having a good experience with feature-based methods from the 1st AnDi Challenge, we tried to use methods with a relatively high level of interpretability, at least as parts of the procedure, also in the just-completed second edition of the competition.

This time, however, the effectiveness of our approach lagged significantly behind the best solutions based on neural networks. We placed 13th in the single-trajectory task and 7th in the ensemble one. From the results it follows that the estimation of the generalized diffusion coefficient $K$ was the weakest point of our approach. Since we used for that purpose an analytical method considered one of the best in its category, this result demonstrates how much of an advantage deep learning methods now have over the analytical approach. Changing the estimation method of $K$ to a gradient-boosted regressor did not help, indicating that the feature set we used may require far-reaching adjustments for the new type of data.

\section*{Acknowledgements}

Jakub Malinowski acknowledges the support of the Dioscuri program
initiated by the Max Planck Society, jointly managed with the
National Science Centre (Poland), and mutually funded by the
Polish Ministry of Science and Higher Education and the
German Federal Ministry of Education and Research.

\section*{Data availability statement}
No new data sets were created or analyzed in this study.

\appendix

\section{Feature set for gradient boosting regression}
\label{app:features}

As already mentioned in Sec.~\ref{subsec:alpha}, we used a method based on gradient boosting to estimate the diffusion exponent  $\alpha$. The set of features constructed for the first AnDi challenge and then further optimized was used to characterize the trajectories. A detailed description of all features can be found in Refs.~\cite{KOW19,LOC20,KOW22}. 

Both the original feature set for the 1st AnDi challenge and its extension are listed in Table~\ref{tab:features}.
Here, we briefly describe them for the convenience of a reader.

\begin{table}[ht!]
    \centering
    \begin{tabular}{c|c}
    \hline\hline
    Original features     & Additional features \\
    \hline\hline
    Anomalous exponent & D'Agostino-Pearson test statistic\\
    \hline
    Diffusion coefficient  & 
    \begin{tabular}{c}Kolmogorov-Smirnov statistic\\ against $\chi^2$ distribution\end{tabular}
    \\
    \hline
    Asymmetry & Noah exponent \\
    \hline
    Efficiency & Moses exponent\\
    \hline
    Empirical velocity autocorrelation function
     &  Joseph exponent
    \\
    \hline
    Fractal dimension & Detrending moving average \\
    \hline
    Maximal excursion &Average moving window characteristics\\
    \hline
    Mean maximal excursion & Maximum standard deviation \\
    \hline
    Mean gaussianity &\\
    \hline
    Mean-squared displacement ratio &\\
    \hline
    Kurtosis &\\
    \hline
    Statistics based on $p$-variation &\\
    \hline
    Straightness &\\
    \hline
    Trappedness & \\
    \hline\hline
    \end{tabular}
    \caption{The features used to characterize the SPT trajectories. The original set of features for the 1st AnDi Challenge have been extended afterwards to improve the performance of the classifier~\cite{KOW22}.\label{tab:features}}
\end{table}

\subsubsection*{Anomalous exponent}

In our set of features, we included 4 estimates for the anomalous exponent $\alpha$, calculated using the following methods:
\begin{itemize}
    \item the standard estimation, based on fitting the empirical time-averaged MSD (TAMSD) from Eq.~(\ref{eq:tamsd}) to the model MSD for anomalous diffusion,
    \begin{equation}
     \widehat{MSD}(n \Delta t) \sim  K_\alpha t^\alpha,\label{eq:msd_th}
    \end{equation}
    \item 3 estimation methods proposed for the trajectories with noise~\cite{LAN18}, under the assumption that the noise is normally distributed with zero mean:
    \begin{itemize}
        \item using estimator
        $$
        \hat{\alpha} = \frac{n_{max} \sum_{n=1}^{n_{max}} \ln (n) \ln (\widehat{MSD}(n\Delta t)) - \sum_{n=1}^{n_{max}} \ln (n)\left( \sum_{n=1}^{n_{max}} \ln (\widehat{MSD}(n\Delta t)) \right)}{n_{max} \sum_{n=1}^{n_{max}} \ln^2(n) - \left(\sum_{n=1}^{n_{max}}\ln(n)\right)^2},
        $$
        with $n_{max}$ equal to 0.1 times the trajectory length, rounded to nearest lower integer (but not less than 4),
        \item  simultaneous fitting of parameters $\hat{K}$, $\hat{\alpha}$ and $\hat{\sigma}$ in the equation 
        $$
            \widehat{MSD}(n\Delta t) = 2 d \hat{K}(n\Delta t)^{\hat{\alpha}}+\hat{\sigma}^2,
        $$
        where  $d$ denotes dimension, $K$ is the  generalized diffusion coefficient and  $\sigma^2$ - the variance of noise,
        \item  simultaneous fitting of parameters $\hat{D}$ and $\hat{\alpha}$  in the equation
        $$
            \widehat{MSD}(n\Delta t) = 2 d \hat{D} (\Delta t)^{\hat{\alpha}}(n^{\hat{\alpha}}-1).
        $$
    \end{itemize}
\end{itemize}

\subsubsection*{Diffusion coefficient}

We used the diffusion coefficient extracted from the fit of the empirical TA-MSD (Eq.~(\ref{eq:tamsd})) to Eq.~(\ref{eq:msd_th}).

\subsubsection*{Asymmetry}
The asymmetry of a trajectory can be derived from the gyration tensor~\cite{SAX93}. For a 2D random walk of $N$ steps, the tensor is given by
\begin{equation}
\mathbf{T} =\left(
\begin{array}{cc}
\frac{1}{N}\sum_{j=1}^N (x_j -\langle x \rangle)^2 & \frac{1}{N}\sum_{j=1}^N (x_j -\langle x \rangle)(y_j -\langle y \rangle) \\ 
\frac{1}{N}\sum_{j=1}^N (x_j -\langle x \rangle)(y_j -\langle y \rangle) & \frac{1}{N}\sum_{j=1}^N (y_j -\langle y \rangle)^2
\end{array} 
\right), \label{eq:tensor}
\end{equation} 
where $\langle x \rangle=(1/N)\sum_{j=1}^N x_j$ is the average of $x$ coordinates over all steps in the random walk. The asymmetry is the defined as
\begin{equation}
A=-\log \left(1 - \frac{(\lambda_1-\lambda_2)^2}{2(\lambda_1+\lambda_2)^2} \right),\label{eq:asymmetry}
\end{equation}
where $\lambda_1$ and $\lambda_2$ are the principle radii of gyration, i.e., the eigenvalues of the tensor $\mathbf{T}$~\cite{HEL07}. {Higher values of $A$ indicate a greater tendency for a preferred direction.} %per https://publications.mpi-cbg.de/Helmuth_2007_4825.pdf} (8)} 

\subsubsection*{Efficiency}
Efficiency $E$ relates the net squared displacement of a particle to the sum of squared step lengths,
\begin{equation}
E = \frac{|X_{N-1}-X_0|^2}{(N-1)\sum_{i=1}^{N-1}|X_i -X_{i-1}|^2}. \label{eq:efficiency}
\end{equation} 

\subsubsection*{Empirical velocity autocorrelation function}

Empirical velocity autocorrelation function \cite{WEB10} for lag $1$ and point $n$ is defined as:
\begin{equation}
    \chi_{n} = \frac{1}{N-1}\sum^{N-2}_{i=0} (X_{i+1+n}-X_{i+n}) (X_{i+1}-X_{i})\label{eq:vac}
\end{equation}
In our model, we used $\chi_n$ for points $n=1$ and $n=2$.

\subsubsection*{Fractal dimension}
Fractal dimension is a measure of the space-filling capacity of a pattern (a trajectory in our case). According to Katz and George~\cite{KAT85}, the fractal dimension of a planar curve may be calculated as 
\begin{equation}
    D_f=\frac{\ln N}{\ln (NdL^{-1})},
\end{equation}
where $L$ is the total length of the trajectory, $N$ is the number of
steps, and $d$ is the largest distance between any two positions. 

\subsubsection*{Maximal excursion}
Maximal excursion of the particle is given by the formula:
\begin{equation}
    ME=\frac{\max_i(X_{i+1}-X_{i})}{X_{N-1}-X_0}.
\end{equation}

\subsubsection*{Mean maximal excursion}

Given the largest distance traveled by a particle,
\begin{equation}
    D_N = \max_i |X_i-X_0|,
\end{equation}
the mean maximal excursion is defined as its standardized value, i.e.:
\begin{equation}
    T_n=\frac{\max_i(|X_i-X_0|)}{\sqrt{\hat{\sigma}^2_N(t_N-t_0)}}.\label{eq:ts}
\end{equation}
Here, $\hat{\sigma}_N$ is a consistent estimator of the standard deviation of $D_N$,
\begin{equation}
    \hat{\sigma}^2_N=\frac{1}{2N\delta t}\sum^{N}_{j=1}||X_j-X_{j-1}||^2_2.
\end{equation}

\subsubsection*{Mean gaussianity}
Gaussianity $g(n)$~\cite{ERN14} checks the Gaussian statistics of increments of a trajectory and is defined as
\begin{equation}
    g(n)=\frac{2\left<r_n^4\right>}{3\left<r_n^2\right>^2},
\end{equation}
where $\left<r_n^k\right>$ denotes the $k$th moment of the trajectory at time lag $n$:
\begin{equation}
    \left<r_n^k\right>=\frac{1}{N-n}\sum^{N-n}_{i=1}|X_{i+n}-X_i|^k.
\end{equation}
Instead of looking at Gaussianities at single time lags, we will include the mean over all lags as one of the features:
\begin{equation}
    \langle g\rangle= \frac{1}{N}\sum^{N}_{i=1}g(n).
\end{equation}

\subsubsection*{Mean-squared displacement ratio}
MSD ratio gives information about the shape of the corresponding MSD curve. We will define it as
\begin{equation}
MSDR(n_1,n_2) = \frac{\langle r_{n_1}^2 \rangle}{\langle r_{n_2}^2 \rangle} - \frac{n_1}{n_2},\label{eq:msdr}
\end{equation}
where $n_1<n_2$. We simply took $n_2=n_1+\Delta t$ and calculate an averaged ratio for every trajectory.

\subsubsection*{Kurtosis}

To calculate kurtosis~\cite{HEL07}, the position vectors $X_i$ are projected onto the dominant eigenvector  $\vec{r}$ of the gyration tensor~(\ref{eq:tensor}),
\begin{equation}
x_i^p = X_i\cdot\vec{r}.
\end{equation}
Kurtosis is then defined as the fourth moment of  $x_i^p$,
\begin{equation}
Kur=\frac{1}{N}\sum_{i=1}^N \frac{(x_i^p -\bar{x}^p)^4}{\sigma^4_{x^p}},\label{eq:kurtosis}
\end{equation}
with $\bar{x}^p$ being the mean projected position and $\sigma_{x^p}$ - the standard deviation of $x^p$.

\subsubsection*{Statistics based on $p$-variation}
The empirical $p$-variation is given by the formula \cite{BUR10,MAG09}:
\begin{equation}
V_m^{(p)}=\sum_{i=0}^{\frac{N}{m}-1}|X_{(i+1)m}-X_{im}|^p.\label{eq:pvar}
\end{equation}
We defined  6 features based on $V_m^{(p)}$:
\begin{itemize}
    \item power $\gamma^p$ fitted to $p$-variation for lags 1 to 5~\cite{BUR10}, 
    \item statistics $P$ used in Ref.~\cite{LOC20}, based on the monotonicity changes of $V_m^{(p)}$ as a function of $m$:
    	$$
		P = \left\{
		\begin{array}{rl}
		0 & \textrm{if $V_m^{(p)}$ does not change the monotonicity}, \\ 
		1 & \textrm{if $V_m^{(p)}$ is convex for the highest $p$ for which it is not monotonous},\\
		-1 & \textrm{if $V_m^{(p)}$ is concave for  the highest $p$ for which it is not monotonous}.
		\end{array} 
		\right.
	$$
\end{itemize}

\subsubsection*{Straightness}
Straightness $S$ measures the average direction change between subsequent steps. It relates the net displacement of a particle to the sum of all step lengths,
\begin{equation}\label{eq:straightness}
S = \frac{|X_{N-1}-X_0|}{\sum_{i=1}^{N-1}|X_i -X_{i-1}|}. 
\end{equation}
{The closer it is to 1, the straighter the trajectory is. In contrast, $S$ close to 0 indicates no preferred direction on the trajectory.}

\subsubsection*{Trappedness}

With trappedness we will refer to the probability that a diffusing particle is trapped in a bounded region with radius $r_0$. A comparison of analytical and Monte Carlo results for confined diffusion allowed Saxton~\cite{SAX93} to estimate this probability with
\begin{equation}
P(K,t,r_0) = 1-\exp\left( 0.2048 -0.25117\left( \frac{Kt}{r_0^2} \right)  \right).\label{eq:trappedness}
\end{equation}
Here, $r_0$ is approximated by half of the maximum distance between any two positions along a given trajectory. $K$ in Eq.~(\ref{eq:trappedness}) is estimated by fitting the first two points of the MSD curve (i.e. it is the so called short-time diffusion coefficient).

\subsubsection*{D’Agostino-Pearson test statistic}

D’Agostino-Pearson $\kappa^2$ test statistic~\cite{AGO71,AGO73} is a goodness-of-fit measure that aims to establish whether or not a given sample comes from a normally distributed sample. It is defined as
\begin{equation}
    \kappa^2 = Z_1(g_1) + Z_2(K),
\end{equation}
where $K$ is the sample kurtosis given by Eq.~(\ref{eq:kurtosis}) and $g_1=m_3/m_2^{3/2}$ is the sample skewness with $m_j$ being the $j$th sample central moment. The transformations $Z_1$ and $Z_2$ should bring the distributions of the skewness and kurtosis as close to the standard normal as possible. Their definitions may be found elsewhere~\cite{AGO71,AGO73}.

\subsubsection*{Kolmogorov-Smirnov statistic against $\chi^2$ distribution}

The Kolmogorov-Smirnov statistic quantifies the distance between the empirical distribution function of the sample $F_n(X)$ and the cumulative distribution function $G_n(X)$ of a reference distribution,
\begin{equation}
   D_n = \sup_X \vert F_n(X) - G_n(X) \vert .
\end{equation}
Here, $n$ is the number of observations (i.e. the length of a trajectory). The value of this statistic for the empirical distribution of squared increments of a trajectory against the sampled $\chi^2$ distribution has been taken as the next feature. The rationale of such choice is that for a Gaussian trajectory the theoretical distribution of squared increments is the mentioned $\chi^2$ distribution.

\subsubsection*{Noah, Moses, and Joseph exponents}

For processes with stationary increments, there are in principle two mechanisms that violate the Gaussian central limit theorem and produce anomalous scaling of MSD~\cite{AGH21}: long-time increment correlations or a flat-tailed increment distribution (in the latter case the second moment is divergent). An anomalous scaling can also be induced by a non-stationary increment distribution. In this case, we deal with the Moses effect. 

All three effects may be quantified by exponents, which will be used as features in our extended set of attributes. Given a stochastic process $X_t$ and the corresponding increment process $\delta_t(\tau)=X_{t+\tau}-X_t$, the Joseph, Moses and Noah exponents are defined as follows.

\begin{enumerate}
    \item Joseph exponent $J$ is estimated from the ensemble average of the rescaled range statistic:
$$E\left[\frac{\max_{1\leq  s\leq  t}[X_s-\frac{s}{t}X_t]-\min_{1\leq  s\leq  t}[X_s-\frac{s}{t}X_t]}{\sigma_t}\right]\sim t^J,$$    
    
$$E\left[\frac{R_t}{S_t}\right]\sim t^J,$$ where $R_t$ is calculated as $R_t=\max_{1\leq  s\leq  t}[X_s-\frac{s}{t}X_t]-\min_{1\leq  s\leq  t}[X_s-\frac{s}{t}X_t]$ and $S_t$ is standard deviation of process $X_t$, {and $E$ is the mathematical expectation.}
\item Moses exponent $M$ is determined from the scaling
of the ensemble probability distribution of the sum of the
absolute value of increments, which can be estimated by
the scaling of the median of the probability distribution
of $Y_t=\sum^{t-1}_{s=0}|\delta_s|$:
$$m[Y_t]\sim t^{M-\frac{1}{2}}$$
\item Noah exponent $L$ is extracted from the scaling of
the ensemble probability distribution of the sum of increment
squares, which can be estimated by the scaling of the median
of the probability distribution of $Z_t=\sum^{t-1}_{s=0}\delta_s^2$:
$$m[Z_t]\sim t^{2L+2M-1}$$
\end{enumerate}

\subsubsection*{Detrending moving average}
The detrending moving average (DMA) statistic is given by
\begin{equation}
    DMA(\tau) = \frac{1}{N-\tau+1} \sum_{i=\tau}^{N} \left( X_i - \overline{X}^{\tau}_i \right)^2,
\end{equation}
for $\tau=1,2,...$, where $\overline{X}^{\tau}_i$ is a moving average of $\tau$ observations, i.e. $\overline{X}^{\tau}_i = \frac{1}{\tau+1} \sum_{j=0}^{\tau} X_{i-j}$~\cite{SIK18,BAL21}.
In our model, we used two values of DMA for each trajectory as input features, namely $DMA(1)$ and $DMA(2)$.

\subsubsection*{Average moving window characteristics}

We added eight features based on the formula
\begin{equation}
    MW = \frac{1}{2(N+1)} \sum_{i=0}^{N} \left| \textrm{sgn} \left(\overline{X}_{i+1}^{(m)}-\overline{X}_{i}^{(m)}\right) - \textrm{sgn}\left(\overline{X}_{i+1}^{(m)}-\overline{X}_{i}^{(m)}\right) \right|,
\end{equation}
where $\overline{X}^{(m)}$ denotes a statistic of the process calculated within the window of length $m$ and $\textrm{sgn}$ is the signum function. In particular, we used the mean and the standard deviation for $\overline{X}$ and calculated $MW$ with windows of  lengths $m=10$ and $m=20$ separately  for $x$ and $y$ coordinates. 

\subsubsection*{Maximum standard deviation}

The idea of the moving window helped us to introduce another two features based on the standard deviation $\sigma_m$ of the process calculated within a window of length $m$. They are given by 
\begin{equation}
    MXM = \frac{\max\left(\sigma_m(t)\right)}{\min\left(\sigma_m(t)\right)}
\end{equation}
and
\begin{equation}
    MXC = \frac{\max\left|\sigma_m(t+1)-\sigma_m(t)\right|}{\sigma},
\end{equation}
where $\sigma$ denotes the overall standard deviation of the sample. We used the window of length $m=3$ and calculated the features for both coordinates separately.

\bibliographystyle{unsrt} 
\bibliography{references}

%\begin{thebibliography}{9}
%\bibitem{andi1}
%P. Kowalek, H. Loch-Olszewska, Ł. Łaszczuk, J. Opała, J. Szwabiński, Boosting the performance of anomalous diffusion classifiers with the proper choice of features, Journal of Physics A 55, 244005 (2022), doi: 10.1088/1751-8121/ac6d2a

%\bibitem{estim}
%Yann Lanoiselée, Grzegorz Sikora, Aleksandra Grzesiek, Denis S. Grebenkov, and Agnieszka Wyłomańska, Optimal parameters for anomalous-diffusion-exponent estimation from noisy data, Phys. Rev. E 98, 062139, doi: 10.1103/PhysRevE.98.062139
%\end{thebibliography}

\clearpage

\appendix

\end{document}